\documentclass[lettersize,journal]{IEEEtran}
\usepackage{amsmath,amsfonts}
\usepackage[ruled,vlined]{algorithm2e}
\usepackage{array}
\usepackage{float}
\usepackage{textcase}
\usepackage[caption=false,font=normalsize,labelfont=sf,textfont=sf]{subfig}
\usepackage{textcomp}
\usepackage{stfloats}
\usepackage{url}
\usepackage{verbatim}
\usepackage{graphicx}
\usepackage{cite}
\usepackage{makecell}
\usepackage{booktabs}
\usepackage{multirow,booktabs,pifont}
\newcommand{\xmark}{\ding{55}}
\newcommand{\cmark}{\ding{51}}
\usepackage{diagbox}
\usepackage{colortbl}
\begin{document}

\title{Audio-visual End-to-end Multi-channel Speech Separation, Dereverberation and Recognition
}

\author{Guinan Li, Jiajun Deng, Mengzhe Geng, Zengrui Jin, Tianzi Wang, Shujie Hu, \\ Mingyu Cui, Helen Meng,~\IEEEmembership{Fellow,~IEEE}, Xunying Liu,~\IEEEmembership{Memeber,~IEEE}
\thanks{Guinan Li, Jiajun Deng, Mengzhe Geng, Zengrui Jin, Tianzi Wang, Shujie Hu, Mingyu Cui are with the Chinese University of Hong Kong, China (email: \{gnli, jjdeng, mzgeng, zrjin, twang, sjhu, mycui\}@se.cuhk.edu.hk)}
\thanks{Helen Meng is with the Chinese University of Hong Kong, China (email:  hmmeng@se.cuhk.edu.hk).}
\thanks{Xunying Liu is with the Chinese University of Hong Kong, China and the corresponding author (email:  xyliu@se.cuhk.edu.hk).}
}

\markboth{Journal of \LaTeX\ Class Files,~Vol.~14, No.~8, August~2021}%
{Shell \MakeLowercase{\textit{et al.}}: A Sample Article Using IEEEtran.cls for IEEE Journals}

\maketitle

\begin{abstract} 
Accurate recognition of cocktail party speech containing overlapping speakers, noise and reverberation remains a highly challenging task to date. Motivated by the invariance of visual modality to acoustic signal corruption, an audio-visual multi-channel speech separation, dereverberation and recognition approach featuring a full incorporation of visual information into all system components is proposed in this paper. The efficacy of the video input is consistently demonstrated in mask-based MVDR speech separation, DNN-WPE or spectral mapping (SpecM) based speech dereverberation front-end and Conformer ASR back-end. Audio-visual integrated front-end architectures performing speech separation and dereverberation in a pipelined or joint fashion via mask-based WPD are investigated. The error cost mismatch between the speech enhancement front-end and ASR back-end components is minimized by end-to-end jointly fine-tuning using either the ASR cost function alone, or its interpolation with the speech enhancement loss. Experiments were conducted on the mixture overlapped and reverberant speech data constructed using simulation or replay of the Oxford LRS2 dataset. The proposed audio-visual multi-channel speech separation, dereverberation and recognition systems consistently outperformed the comparable audio-only baseline by 9.1\% and 6.2\% absolute (41.7\% and 36.0\% relative) word error rate (WER) reductions. Consistent speech enhancement improvements were also obtained on PESQ, STOI and SRMR scores\footnote{Enhanced audio examples for demonstration purposes are available in https://liguinan.github.io/AV-E2E-MC-ASR}.
\end{abstract}

\section{Introduction}
\IEEEPARstart{D}{espite} the rapid progress of automatic speech recognition (ASR) in the past few decades, accurate recognition of cocktail party  speech \cite{mcdermott2009cocktail, qian2018past} remains a highly challenging task to date. 
Its difficulty can be attributed to multiple sources of interference including overlapping speakers, background noise and room reverberation. These lead to a large mismatch between the resulting mixture speech and clean signals.

To this end, microphone arrays play a key role in state-of-the-art speech enhancement and recognition systems designed for cocktail party overlapped speech and far-field scenarios \cite{yoshioka2015ntt, chang2019mimo, haeb2020far}. 
The required array beamforming techniques used to perform multi-channel signal integration are normally implemented as either time or frequency domain filters.
These are represented by time domain delay and sum \cite{anguera2007acoustic}, frequency domain minimum variance distortionless response (MVDR) \cite{souden2009optimal, van2002optimum} and generalized eigenvalue (GEV) \cite{warsitz2007blind} based multi-channel integration approaches.
Earlier generations of mixed speech separation and recognition systems featuring conventional multi-channel array beamforming techniques typically used a pipelined system architecture. 
It contains separately constructed speech enhancement front-end modules designed to perform speech separation, dereverberation as well as denoising tasks, and speech recognition back-end components.

With the wider application of deep neural networks (DNNs) based speech technologies, microphone array beamforming techniques have also evolved into a rich variety of neural network based designs in recent few years. 
These include: 
a) neural time-frequency (TF) masking approaches \cite{bahmaninezhad2019,chen2019multi, gu2020enhancing} used to predict spectral mask labels for a reference channel that specify whether a particular TF spectrum point is dominated by the target speaker or interfering sources to facilitate speech separation; 
b) neural Filter and Sum approaches directly estimating the beamforming filter parameters in either time domain \cite{sainath2017multichannel} or frequency domain \cite{xiao2016deep} to produce the separated outputs; 
and c) mask-based MVDR \cite{ yoshioka2018recognizing, chang2019mimo, yoshioka2018multi, higuchi2018frame, kubo2019mask, xu2019joint}, and mask-based GEV \cite{heymann2017beamnet, drude2018integrating} approaches utilizing DNN estimated TF masks to compute target speaker and noise specific speech power spectral density (PSD) matrices and to obtain the beamforming filter parameters, while alleviating the need of explicit direction of arrival (DOA) estimation.

In many practical applications, reverberation presents a further challenge which can lead to severe speech recognition performance degradation \cite{kinoshita2017neural, heymann2019joint} when such systems are trained on anechoic and non-reverberant data. 
Classical solutions to the resulting dereverberation problem represented by, for example, weighted prediction error (WPE) \cite{nakatani2010speech}, require the estimation of a time delayed linear filter.
In recent years, there has been a similar trend of conventional speech dereverberation approaches \cite{furuya2007,nakatani2008blind, nakatani2010speech, huang2022kronecker} such as WPE evolving into their current DNN based variants. These include: 
a) the DNN-WPE \cite{kinoshita2017neural, heymann2019joint} method, which uses neural network estimated target signal PSD matrices in place of those traditionally obtained using maximum likelihood estimation trained complex value Gaussian Mixture Models \cite{nakatani2010speech} in the dereverberation filter estimation; 
and b) complex spectral masking \cite{williamson2017time, fu2022uformer} and spectral mapping \cite{wang2020multi, zhao2020monaural} learning a transformation between reverberant and anechoic data.

End-to-end all neural microphone array based speech enhancement and recognition systems present a comprehensive and overarching solution to the cocktail party speech problem by simultaneously performing speech separation, denoising and dereverberation. However, efforts on developing such systems are confronted by a number of key research challenges.

\textbf{1) Full incorporation of video modality:} Motivated by the bimodal nature of human speech perception and the invariance of visual information to extrinsic acoustic corruption, there has been a long history of developing audio-visual speech enhancement \cite{michelsanti2021overview, afouras2018conversation, wu2019time, ephrat2018looking, gu2020multi, sato2021multimodal, ochiai2019multimodal, xu2020neural, wu2022time, li2020deep, lu2019audio, chung2020facefilter,  wang2020robust, iuzzolino2020av, morrone2019face, gabbay2018seeing, tan2020audio, michelsanti2019training} and recognition \cite{afouras2018deep, noda2015audio, mroueh2015deep, zhang2019robust, wu2021audio, petridis2018audio, tao2018gating, su2019cross, abdelaziz2017comparing, shi2022robust, ma2021end, petridis2018end, braga2020end, yu2021fusing, li2019improving, zhou2019modality, rose22_interspeech} techniques. 
When processing the cocktail mixed speech, a holistic, consistent incorporation of visual information in all components of the entire system (speech separation, dereverberation and recognition) is preferred. 
In contrast, among existing researches, video information has mainly been partially incorporated into: 
a) the speech enhancement (separation and/or dereverberation) front-end \cite{afouras2018conversation, wu2019time, ephrat2018looking, gu2020multi, sato2021multimodal, ochiai2019multimodal, xu2020neural, wu2022time, li2020deep, lu2019audio, chung2020facefilter, wang2020robust, iuzzolino2020av, morrone2019face, gabbay2018seeing, tan2020audio, michelsanti2019training} alone; 
or b) the speech recognition back-end \cite{afouras2018deep, noda2015audio, mroueh2015deep, zhang2019robust, wu2021audio, petridis2018audio, tao2018gating, su2019cross, abdelaziz2017comparing, shi2022robust, ma2021end, petridis2018end, braga2020end, yu2021fusing, zhou2019modality, li2019improving, rose22_interspeech} only. 
More recent works used video information in both the multi-channel speech separation and ASR \cite{yu2021audio}, but not in speech dereverberation.

\textbf{2) Integration between speech separation and dereverberation modules:} Surface reflection of speech signals in reverberant environments distorts the DOA or TF-mask estimation for the target speaker. 
At the same time, interfering sound sources also impact the dereverberation filter estimation.  Hence, a suitable form of integration between the speech separation and dereverberation techniques is required within the speech enhancement front-end sub-system. 
Possible integration solutions include: 
a) a pipelined architecture within which the speech separation and dereverberation components are sequentially connected in any order such as the previous researches in \cite{drude2018integrating, tan2020audio, li2022audio}; 
or b) a single architecture where both these two enhancement functions are implemented, for example, using weighted power minimization distortionless response (WPD) \cite{nakatani2019unified, nakatani2019simultaneous, 2020Jointly} and the related DNN TF-mask based WPD \cite{zhang2022end, ni2021wpd++} approaches. To date, such integration problem has only been investigated for audio-only speech enhancement \cite{drude2018integrating, delcroix2015strategies, nakatani2020dnn, boeddeker2020jointly, nakatani2019unified, 2020Jointly, ni2021wpd++, nakatani2021blind, nakatani2019simultaneous, zhang2022end}, but has not been studied for audio-visual speech separation and dereverberation. 

\textbf{3) Joint optimization of audio-visual speech enhancement front-end and recognition back-end:} Conventional non-DNN based speech enhancement front-end models are often separately constructed and cannot be easily integrated with the ASR back-end. 
The wide application of deep learning approaches for speech enhancement and recognition components allows them to be more tightly integrated and consistently optimized in an end-to-end manner. 
An improved trade-off between the speech enhancement front-end loss function and ASR accuracy can then be obtained, for example, using multi-task learning \cite{von2020multi, yu2021audio, wang2022sjtu}. 
To date, such joint speech enhancement front-end and ASR back-end optimization has been only conducted among: a) audio-only speech enhancement and recognition systems using no video input \cite{von2020multi, subramanian2020far, xu2019joint, shao2022multi, heymann2019joint, kumar2022end, zhang2022end}; or b) audio-visual speech separation and recognition tasks only while not considering speech dereverberation \cite{yu2021audio, wang2022sjtu}. 
Hence, there is a pressing need to derive suitable joint optimization methods for a complete audio-visual multi-channel speech separation, dereverberation and recognition system. 

In order to address the above issues, an audio-visual multi-channel speech separation, dereverberation and recognition approach featuring a full incorporation of visual information into all three components of the entire system is proposed in this paper.
The efficacy of the video input is consistently demonstrated when being used in the mask-based MVDR speech separation, DNN-WPE or spectral mapping (SpecM) based speech dereverberation front-end and Conformer encoder-decoder based ASR back-end components. 
Both the pipelined integration methods using either a) a serial connection of the audio-visual speech separation component with the following dereverberation module; or  b) audio-visual speech dereverberation followed by separation; and c) joint speech separation and dereverberation via audio-visual mask-based WPD are investigated. 
In order to reduce the error cost mismatch between the speech enhancement front-end and ASR back-end components, they are jointly fine-tuned using either only the Conformer ASR cost function (CTC plus Attention) \cite{watanabe2017hybrid}, or the ASR cost function interpolated with the speech enhancement loss based on mean square error (MSE) and scale-invariant signal to noise ratio (SISNR). 

Experiments conducted on the mixture overlapped and reverberant speech data constructed using either simulation or replay of the benchmark Oxford LRS2 dataset \cite{son2017lip} suggest:

1) The proposed audio-visual multi-channel speech separation, dereverberation and recognition systems consistently outperformed the comparable audio-only baseline systems by {\bf 9.1\% and 6.2\% absolute  (41.7\% and 36.0\% relative)} word error rate (WER) reductions on the LRS2 simulated and replayed evaluation datasets, respectively. Consistent improvements of perceptual evaluation of speech quality (PESQ) \cite{recommendation2001perceptual}, short-time objective intelligibility (STOI) \cite{taal2011algorithm} and speech to reverberation modulation energy ratio (SRMR) \cite{falk2010non} scores were also obtained. 

2) In particular, when compared with audio-only dereverberation, incorporating visual information into the DNN-WPE or SpecM based dereverberation module produced consistent improvements of PESQ, STOI and SRMR scores and a statistically significant\footnote{Matched pairs sentence-segment word error (MAPSSWE) based statistical significance test \cite{Gillick1989SomeSI} was performed at a significance level $\alpha$=0.05.} WER reduction by up to {\bf 1.9\% absolute (5.9\% relative)}, irrespective of the form of integration between speech separation and dereverberation components. 

3) Among different architectures to integrate the speech separation and dereverberation components within the front-end, a pipelined, full audio-visual configuration performing DNN-WPE based speech dereverberation followed by mask-based MVDR speech separation using video input in both stages produced the best overall speech enhancement and recognition performance. 

4) Consistent WER reductions and improvements on speech enhancement metric scores were also obtained after joint fine-tuning the entire audio-visual speech separation, dereverberation and recognition system in a fully end-to-end manner. 

The main contributions of this paper are summarized below:

1) To the best of our knowledge, this paper presents the first use of a complete audio-visual multi-channel speech separation, dereverberation and recognition system architecture featuring a full incorporation of visual information into all three stages. In contrast, prior researches incorporate visual modality in either only the speech enhancement front-end \cite{afouras2018conversation, wu2019time, ephrat2018looking, gu2020multi, sato2021multimodal, ochiai2019multimodal, xu2020neural, wu2022time, li2020deep, lu2019audio, chung2020facefilter,  wang2020robust, iuzzolino2020av, morrone2019face, gabbay2018seeing, tan2020audio, michelsanti2019training}, ASR back-end \cite{afouras2018deep, noda2015audio, mroueh2015deep, zhang2019robust, wu2021audio, petridis2018audio, tao2018gating, su2019cross, abdelaziz2017comparing, shi2022robust, ma2021end, petridis2018end, braga2020end, yu2021fusing, zhou2019modality,li2019improving, rose22_interspeech}, or both the multi-channel speech separation and recognition stages \cite{yu2021audio} but excluding the dereverberation component.

2) This paper presents a more complete investigation of the advantages of audio-visual dereverberation approaches versus audio-only dereverberation methods based on  DNN-WPE and SpecM. In contrast, similar prior studies \cite{tan2020audio} were conducted only in the context of SpecM based dereverberation. 

3) To the best of our knowledge, this is the first work that systematically investigates the suitable form of integration between the full audio-visual speech separation and dereverberation modules within the speech enhancement front-end. In contrast, similar studies in previous researches were only conducted for audio-only speech enhancement \cite{zhang2022end}.

4) This paper presents the first research to demonstrate that performing an end-to-end joint optimization is useful for training a complete audio-visual multi-channel speech separation, dereverberation and recognition system. In contrast, related prior studies were conducted only in the context of audio-only speech enhancement and recognition \cite{zhang2022end}. 

We hope these findings above will provide valuable insights for the practical development of state-of-the-art audio-visual speech separation, dereverberation and recognition systems for cocktail party and far-field scenarios.

The rest of the paper is organized as follows. Audio-visual multi-channel speech separation is reviewed in Section II. Section III presents audio-visual multi-channel speech dereverberation. Integrated audio-visual speech separation and dereverberation approaches are proposed in Section IV. Section V presents the audio-visual Conformer ASR back-end component and its joint fine-tuning with the speech enhancement front-end. Experimental data setup and results are presented in Section VI and VII, respectively. Section VIII draws the conclusion and discusses future research directions.

\begin{figure*}[htbp]
    \centering
    \setlength{\abovecaptionskip}{0pt plus 1pt minus 3pt}
    \includegraphics[scale=0.53]{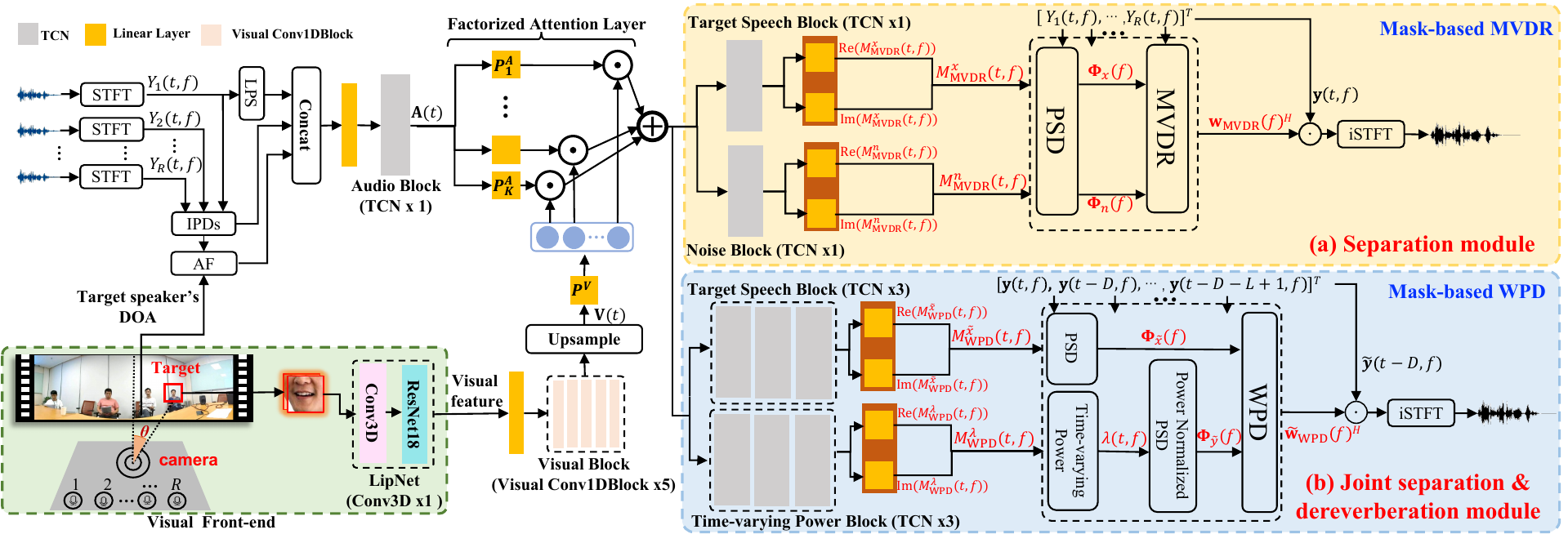}
    \caption{Audio-visual multi-channel \textbf{speech separation} using mask-based MVDR approach (a), and \textbf{joint speech separation \& dereverberation module} using mask-based WPD in (b). Both use the same audio-visual embeddings (left part of the figure) for their complex masks estimation.
     $Y_r(t,f) \in \mathbb{C} $ is the $r$-th channel's complex spectrum of mixture speech among $R$ microphone channels. $\mathbf{V}(t)$ and $\mathbf{A}(t)$ denote the audio and visual embeddings at frame index $t$, respectively. The internal structural details of the TCN block and Visual Conv1DBlock are shown in Fig. \ref{fig:tcn_visual_block_module}.  
    The MVDR filter $\mathbf{w}_{\text{\tiny{MVDR}}}(f) \in \mathbb{C}^{R}$ is estimated using the target speech and noise PSD matrices $\boldsymbol{\Phi}_x(f) \in \mathbb{C}^{R\times R}$ and $\boldsymbol{\Phi}_n(f) \in \mathbb{C}^{R\times R}$ with their respective complex TF masks $M^x_{\text{\tiny{MVDR}}}(t, f) \in \mathbb{C}$ and $M^n_{\text{\tiny{MVDR}}}(t, f) \in \mathbb{C}$. 
    The WPD filter $\tilde{\mathbf{w}}_{\text{\tiny{WPD}}}(f) \in \mathbb{C}^{(L+1)R}$ is estimated using the target speaker and power normalized spatial-temporal PSD matrices $\boldsymbol{\Phi}_{\tilde{x}}(f) \in \mathbb{C}^{(L+1)R \times (L+1)R}$ and $\boldsymbol{\Phi}_{\tilde{y}}(f) \in \mathbb{C}^{(L+1)R \times (L+1)R}$ with their respective complex TF masks $M^{\tilde{x}}_{\text{\tiny{WPD}}}(t, f) \in \mathbb{C}$ and $M^{\lambda}_{\text{\tiny{WPD}}}(t, f) \in \mathbb{C}$. 
     $\text{Re}(\cdot)$ and $\text{Im}(\cdot)$ denote the real and imaginary parts operators. $D$ is the prediction delay parameter and $L$ is the number of filter taps.
    }
    \label{fig:mvdr_wpd_module}
    \vspace{-0.2cm}
\end{figure*}

\vspace{-0.2cm}
\section{Audio-Visual Multi-Channel Speech Separation} \label{section:sep}
In this section, the multi-channel far-field speech signal model is reviewed first, before the introduction of the audio-visual multi-channel mask-based MVDR approach for speech separation is presented. 

\vspace{-0.3cm}
\subsection{Multi-channel Far-field Signal Model}
In the far-field scenarios, the short-time Fourier transform (STFT) spectrum of the received multi-channel speech signal $\mathbf{y}(t, f) \in \mathbb{C}^{R}$ recorded by a microphone array consisting of $R$ channels can be modeled as:
\begin{align}
\mathbf{y}(t, f) = \mathbf{x}(t, f) + \mathbf{n}(t, f) = \mathbf{g}(f)S(t, f) + \mathbf{n}(t, f), \label{equation:signal_model} 
\end{align}
where $t$ and $f$ denote the indices of time and frequency bins, respectively. $\mathbf{x}(t, f) \in \mathbb{C}^{R}$ is a complex vector containing {the clean speech signals received by the array channels}.  $\mathbf{n}(t, f) \in \mathbb{C}^{R}$ represents either the interfering speaker’s speech or additive background noise alone, or a combination of both. $\mathbf{g}(f) \in \mathbb{C}^{R}$ denotes the array steering vector and $S(t, f)$ is the STFT spectrum of the target speaker's clean speech. 

\vspace{-0.2cm}
\subsection{Mask-based MVDR}
Classic acoustic beamforming approaches \cite{souden2009optimal, warsitz2007blind, van2002optimum} are designed to capture the speech from the target speaker’s direction while attenuating the interfering sounds coming from other locations. This is realized by setting, or ``steering", the beamforming filter parameters to the target direction. Taking the MVDR beamformer as an example, a linear filter $\mathbf{w}_{\text{\tiny{MVDR}}}(f) \in \mathbb{C}^{R}$ is applied to the multi-channel mixture speech spectrum $\mathbf{y}(t, f)$ to produce the filtered output ${\hat S}_{\text{\tiny{MVDR}}}(t,f)$ as:
\begin{align} 
\hat{S}_{\text{\tiny{MVDR}}}(t, f) & =\mathbf{w}_{\text{\tiny{MVDR}}}(f)^H \mathbf{y}(t, f),\label{equation:linear_filtering_1}\\
& = \underbrace{\mathbf{w}_{\text{\tiny{MVDR}}}(f)^H \mathbf{x}(t, f)}_{\text{target speech
 component}} + \underbrace{\mathbf{w}_{\text{\tiny{MVDR}}}(f)^H\mathbf{n}(t, f)}_{\text{residual noise}},
 \label{equation:linear_filtering_2}
\end{align}
where $(\cdot)^{H}$ denotes the conjugate transpose operator. 

The MVDR beamformer is designed to minimize the residual noise output while imposing a distortionless constraint on the target speech \cite{souden2009optimal}, which can be formulated as
\begin{align} \label{equation:mvdr_constraint}
& \min _{\mathbf{w}_{\text{\tiny{MVDR}}}(f)} \sum_t \left|   \mathbf{w}_{\text{\tiny{MVDR}}}(f)^{H} \mathbf{n}(t, f) \right|^2,\\
& \text { subject to : } \sum_t \left|\left(\mathbf{u}_r-\mathbf{w}_{\text{\tiny{MVDR}}}(f)\right)^{H} \mathbf{x}(t, f) \right|^2=0,
\end{align}
where $\mathbf{u}_r=[0,0, \ldots, 1, \ldots, 0]^T \in \mathbb{R}^{R}$ is a one-hot reference vector where its $r$-th component equals to one. $(\cdot)^{T}$ denotes the transpose operator. Without loss of generality, we select the first channel, i.e., $r=1$ as the reference channel among the $R$ channels throughout this paper.

The distortionless constraint in the above optimization problem is equivalent to $\mathbf{w}_{\text{\tiny{MVDR}}}(f)^{H} \mathbf{g}(f)=1$, which can be interpreted as maintaining the energy along the target direction. 
The MVDR beamforming filter is estimated as 
\begin{align}
{\small \mathbf{w}_{\text{\tiny{MVDR}}}(f)\!\!=\!\!\frac{\boldsymbol{\Phi}_n(f)^{-1} \mathbf{g}(f)}{\mathbf{g}(f)^{H} \boldsymbol{\Phi}_n(f)^{-1} \mathbf{g}(f)}=\frac{\boldsymbol{\Phi}_n(f)^{-1} \boldsymbol{\Phi}_x(f)} {\operatorname{tr}\left(\boldsymbol{\Phi}_n(f)^{-1} \boldsymbol{\Phi}_x(f)\right)} \mathbf{u}_{r}},\label{equation:mvdr_filter_2}
\end{align}
where the target speaker and noise specific power spectral density (PSD) matrices 
\begin{align}
& {\small \boldsymbol{\Phi}_x(f) \!=\!\frac{\sum_{t}\left(M^x_{\text{\tiny{MVDR}}}(t, f) \mathbf{y}(t, f)\right)\left(M^x_{\text{\tiny{MVDR}}}(t, f) \mathbf{y}(t, f)\right)^{H}}{\sum_{t} M^x_{\text{\tiny{MVDR}}}(t, f) \left(M^x_{\text{\tiny{MVDR}}}(t, f)\right)^{*}}}, \label{equation:mvdr_mask_PSD_1} \\
 & {\small \boldsymbol{\Phi}_n(f) \!=\!\frac{\sum_{t}\left(M^n_{\text{\tiny{MVDR}}}(t, f) \mathbf{y}(t, f)\right)\left(M^n_{\text{\tiny{MVDR}}}(t, f)\mathbf{y}(t, f)\right)^{H}}{\sum_{t} M^n_{\text{\tiny{MVDR}}}(t, f)\left(M^n_{\text{\tiny{MVDR}}}(t, f)\right)^{*}}}, \label{equation:mvdr_mask_PSD_2}
\end{align} 
are computed using DNN predicted complex TF masks $M^x_{\text{\tiny{MVDR}}}(t, f)\in \mathbb{C}$ and $M^n_{\text{\tiny{MVDR}}}(t, f) \in \mathbb{C}$\cite{xu2019joint, yu2021audio}. $\operatorname{tr}(\cdot)$ denotes the trace operator. $(\cdot)^{*}$ is complex conjugate operator.

\vspace{-0.4cm}
\subsection{Audio Modality} \label{subsection:audio_modality}
As is illustrated in the top left corner of Fig. \ref{fig:mvdr_wpd_module}, three types of audio features including the complex STFT spectrum of all the microphone array channels, the inter-microphone phase differences (IPDs) \cite{yoshioka2018recognizing} and location-guided angle feature (AF) \cite{ chen2018multi} are adopted as the audio inputs. IPDs features are used to capture the relative phase difference between different microphone channels and provide additional spatial cues for mask-based multi-channel speech separation. 
Angle features that are based on the approximated DOA of the target speaker\footnote{The target speaker is located using a 180-degree wide-angle camera to track the speaker’s face.
The camera approximated DOA of target speaker is only used in AF features.} are also incorporated to provide further spatial filtering constraints. 
In this work, the approximated DOA of the target speaker is obtained by tracking the speaker’s face from a $180^{\circ}$ wide-angle camera (Fig. \ref{fig:mvdr_wpd_module}, bottom left corner).

\begin{figure}[htbp]
    \centering
    \setlength{\abovecaptionskip}{0pt plus 1pt minus 3pt}
    \includegraphics[scale=0.35]{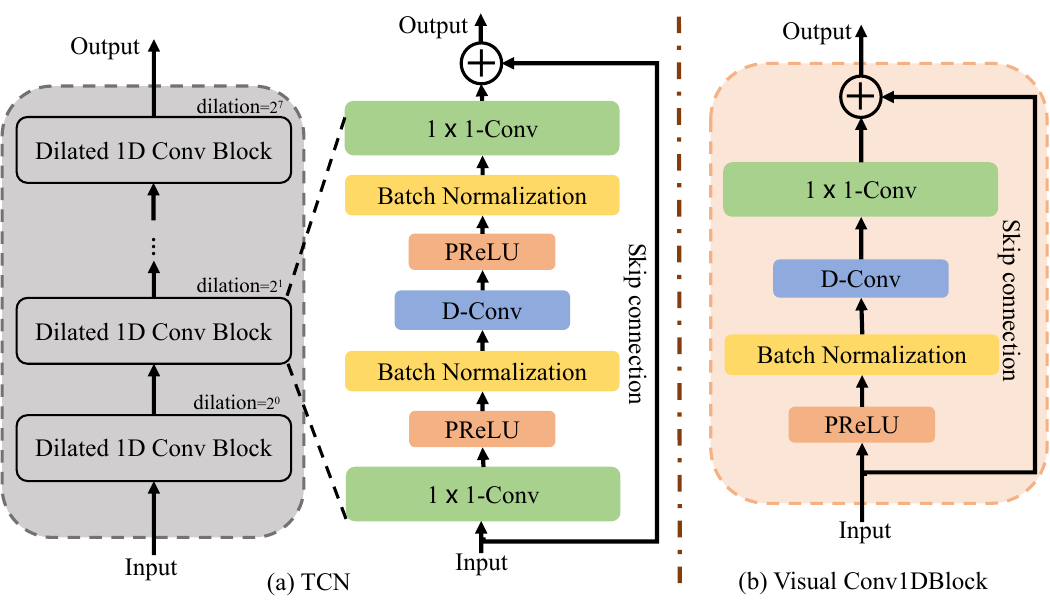}
    \caption{Illustration of the architectures of: (a) the temporal convolutional network (TCN) Block. Each dilated 1-D ConvBlock consists of a $1 \times 1$ convolutional layer, a depth-wise separable convolution layer (D-Conv) \cite{chollet2017xception}, with PReLU \cite{he2015delving} activation function and batch normalization added between each two convolution layers. Skip connections are added in each dilated 1-D ConvBlock; and (b) Visual Conv1DBlock which consists of a PReLU \cite{he2015delving} activation function, batch normalization, a depth-wise separable convolution layer (D-Conv) \cite{chollet2017xception} and a $1 \times 1$ convolutional layer with skip connection.}
    \label{fig:tcn_visual_block_module}
    \vspace{-0.2cm}
\end{figure}

Following prior researches on audio-visual multi-channel speech separation \cite{yu2021audio, li2022audio}, the temporal convolutional network architecture (TCN) \cite{luo2019conv}, which uses a long reception field to capture more sufficient contextual information, is used in our separation system. As shown in the left of Fig. \ref{fig:tcn_visual_block_module}, each TCN block is stacked by 8 Dilated 1-D ConvBlock with exponentially increased dilation factors $2^0, 2^1, \ldots ., 2^7$. As shown in the top left corner of Fig. \ref{fig:mvdr_wpd_module}, the log-power spectrum (LPS) features of the reference microphone channel are concatenated with the IPDs and AF features before being fed into a single TCN module based Audio Block to compute the audio embeddings $\mathbf{A} \in \mathbb{R}^{F_a \times T_a}$, where $F_a$ is the dimension of audio embeddings and $T_a$ is the number of audio frames.

\vspace{-0.4cm}
\subsection{Visual Modality} \label{subsection:visual_modality}
The lip region of a target speaker obtained via face tracking is fed into a LipNet \cite{afouras2018deeplip} which consists of a 3D convolutional layer (Fig. \ref{fig:mvdr_wpd_module}, bottom left, in pink) and an 18-layer ResNet \cite{he2016deep} (Fig. \ref{fig:mvdr_wpd_module}, bottom left, in light turquoise), to extract the visual features from the target speaker’s lip movements.
Before fusing the visual features with the audio embeddings to improve the TF masks estimation, the visual features are firstly fed into the linear layer followed by the Visual Block containing five Visual Conv1DBlocks (Fig. \ref{fig:mvdr_wpd_module}, bottom, in light brown, the detailed network architecture is illustrated in the right of Fig. \ref{fig:tcn_visual_block_module}), and then the output of Visual Block is up-sampled to be time synchronised with the audio frames via linear interpolation to compute the visual embeddings $\mathbf{V} \in \mathbb{R}^{F_v \times T_a}$, where $F_v$ is the dimension of visual embeddings. In this work, the LipNet model is pretrained on the lipreading task as described in \cite{afouras2018deeplip}.

\vspace{-0.4cm}
\subsection{Modality Fusion} \label{subsection:modality_fusion}
In order to effectively integrate the audio and visual embeddings, a factorized attention-based modality fusion method \cite{yu2021audio, li2022audio} is utilized in the audio-visual speech separation module. 
As shown in Fig. \ref{fig:mvdr_wpd_module} (middle up), the acoustic embeddings at frame index $t$ denoted by $\mathbf{A}(t)$ are first factorized into $K$ acoustic subspace vectors $\left[\mathbf{e}_{1}^{a}(t), \mathbf{e}_{2}^{a}(t),\ldots, \mathbf{e}_{K}^{a}(t)\right]$ by a series of parallel linear transformation $\mathbf{P}_k^a \in \mathbb{R}^{F_a \times F_a}$. The visual embeddings at frame index $t$ named by $\mathbf{V}(t)$ is mapped into a $K$ dimensional vector $\mathbf{e}^{v}(t)=\left[e_{1}^{v}(t), e_{2}^{v}(t), \ldots,e_{K}^{v}(t)\right]^T$ by projection matrix $\mathbf{P}^v \in \mathbb{R}^{K \times F_v}$ as
\begin{align}
& {\left[\mathbf{e}_{1}^{a}(t), \mathbf{e}_{2}^{a}(t), \ldots, \mathbf{e}_{K}^{a}(t)\right]= \left[\mathbf{P}_1^a, \mathbf{P}_2^a, \ldots, \mathbf{P}_K^a\right]\mathbf{A}(t)}, \\
& \mathbf{e}^{v}(t)=\operatorname{Softmax}\left( \mathbf{P}^v \mathbf{V}(t)\right), 
\end{align}
Then the fused audio-visual embeddings $\mathbf{AV}(t) \in \mathbb{R}^{F_a}$ are 
\begin{equation}
\mathbf{A V}(t)=\boldsymbol{\sigma}\left(\sum_{k=1}^K e_k^v(t) \mathbf{e}_k^a(t)\right),
\end{equation}
 where $ \boldsymbol{\sigma(\cdot)}$ is the sigmoid function. 

 {The above audio-visual embeddings are fed into both the Target Speech Block and Noise Block (Fig. \ref{fig:mvdr_wpd_module}, center), before their respective outputs being further fed into the corresponding linear layers (Fig. \ref{fig:mvdr_wpd_module}, top right, yellow blocks) to estimate the complex TF masks $M^x_{\text{\tiny{MVDR}}}(t, f)\in \mathbb{C}$ and $M^n_{\text{\tiny{MVDR}}}(t, f) \in \mathbb{C}$ required by the target speech and noise PSD matrices in Eqns. (\ref{equation:mvdr_mask_PSD_1}) and (\ref{equation:mvdr_mask_PSD_2})
for MVDR filter estimation. After  MVDR filtering, the separated target speech spectrum is inverse STFT (iSTFT) transformed to produce the corresponding waveform.}

\vspace{-0.4cm}
\subsection{Separation Network Training Cost Function}
Following the prior researches\cite{tan2020audio, yu2021audio, li2022audio, gu2020multi}, the mask-MVDR based multi-channel speech separation network is trained to maximize the SISNR metric, unless further joint fine-tuning with the back-end ASR error loss later presented in Section \ref{section:comformer_asr} is performed. 

\begin{figure*}[htbp]
    \centering
    \setlength{\abovecaptionskip}{0pt plus 1pt minus 3pt}
    \includegraphics[scale=0.53]{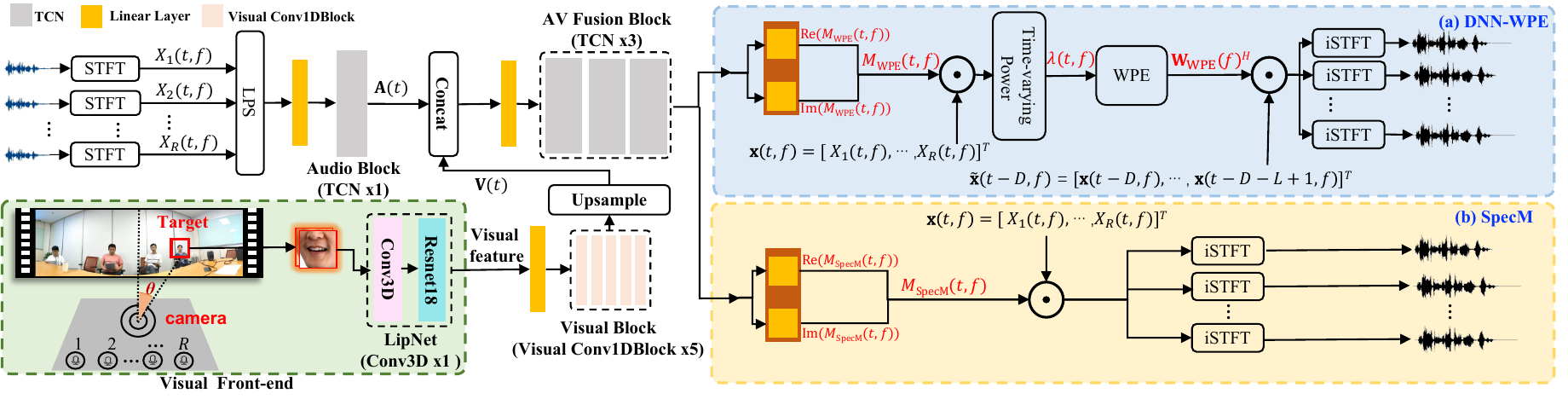}
    \caption{Illustration of audio-visual multi-channel speech dereverberation networks based on the \textbf{(a) {DNN-WPE}} or \textbf{(b) SpecM} approaches of Sections \ref{subsection:DNN-WPE} and \ref{subsection:SpecM} respectively.
    $X_r(t, f) \in \mathbb{C}$ is the $r$-th channel's complex spectrum of reverberant speech among R microphone channels. $\mathbf{V}(t)$ and $\mathbf{A}(t)$ denote the audio and visual embeddings at frame index $t$, in common with Fig.\ref{fig:mvdr_wpd_module}.
    During WPE filter estimation, the signal variance $\lambda(t,f)$ is obtained using DNN predicted TF complex mask $M_{\text{\tiny{WPE}}}(t,f) \in \mathbb{C}$. $\mathbf{x}(t, f) \in \mathbb{C}^R$ is the input multi-channel reverberant speech signal. $D$ denotes the prediction delay parameter and $L$ is the number of filter taps. 
    $M_{\text{\tiny{SpecM}}}(t,f) \in \mathbb{C} $ denotes the complex TF mask for SpecM based dereverberation.}
    \label{fig:dervb_module}
    \vspace{-0.5cm}
\end{figure*}

\vspace{-0.2cm}
\section{Audio-Visual Multi-Channel Speech Dereverberation}\label{section:dervb}
In this section, the multi-channel far-field signal model is reformulated with additional reverberation.
Audio-visual multi-channel speech dereverberation approaches based on audio-visual DNN-WPE and SpecM are then proposed.
The incorporation of the video features and its fusion with audio modality in both methods are also presented.

\vspace{-0.4cm}
\subsection{Multi-channel Far-field Signal Model with Reverberation}
In reverberant conditions, the target speech signal $\mathbf{x}(t, f)$ of Eqn. (\ref{equation:signal_model}) is further decomposed into two parts.
The first part consists of the direct signal and early reflections, referred to as the desired signal $\mathbf{d}(t, f) \in \mathbb{C}^{R}$, while the other contains the late reverberation $\mathbf{r}(t, f) \in \mathbb{C}^{R} $. This is given by
\vspace{-0.2cm}
\begin{align} \label{equation:signal_model_reverb}
\mathbf{x}(t, f) \!\!=  \!\!\! \underbrace{\sum_{\tau=0}^{D-1} \mathbf{a}(\tau, f) S(t\!\!-\!\tau, f)}_{\mathbf{d}(t, f) } \! + \!\!\!\! \underbrace{\sum_{\tau=D}^{D+L-1} \mathbf{a}(\tau, f) S(t\!\!-\!\tau, f)}_{\mathbf{r}(t, f)}
\end{align}
where $D$ denotes the prediction delay parameter and $L$ is the number of filter taps. $\mathbf{a}(\tau, f) \in \mathbb{C}^{R}$ is the room reverberant transfer function from a given speaker to all microphones for $\tau \in \left\{ 0, 1, \ldots, D+L-1 \right\}$. 
The dereverberation process requires the desired signal $\mathbf{d}(t, f)$ to be preserved to enhance speech intelligibility and improve ASR performance, while the late reverberation $\mathbf{r}(t, f)$ to be eliminated \cite{nakatani2010speech}.

\subsection{DNN-WPE Based Dereverberation} \label{subsection:DNN-WPE}
 In conventional WPE \cite{nakatani2010speech}, 
 the dereverberated signal $\hat{\mathbf{d}}(t, f)$ can be obtained by applying the WPE filter $\mathbf{W}_{\text{\tiny{WPE}}}(f) \in \mathbb{C}^{LR \times R}$ to the reverberant multi-channel signal as follows:  
\begin{align} \label{equation:dnn_wpe_filtering}
\hat{\mathbf{d}}(t, f) = \mathbf{x}(t, f) - \mathbf{W}_{\text{\tiny{WPE}}}(f)^H \tilde{\mathbf{x}}(t-D, f),
\end{align}
where $\tilde{\mathbf{x}}(t\!-\!D, f) \!\! = \!\! \left[\mathbf{x}(t\!-\!D, f)^{T}, \ldots, \mathbf{x}(t\!-\!\!D\!-\!L\!+\!1, f)^{T}\right]^{T} \in \mathbb{C}^{LR}$
is the time-delayed reverberant speech spectrum vector.

The required WPE filter coefficients are traditionally estimated using maximum likelihood estimation \cite{nakatani2010speech}.
It is assumed that the desired signal at each microphone follows a time-varying complex Gaussian distribution with a mean of zero and a time-varying variance $\lambda(t, f)$, which corresponds to the power of the desired signal. Minimizing the average power of the frame prediction errors weighted by $\lambda^{-1}(t, f)$,
\begin{equation} 
{\small \min_{\left\{\mathbf{W}_{\text{\tiny{WPE}}}(f) ,\lambda(t, f)\right\}} \sum_t \frac{\left\| \mathbf{x}(t, f) \!-\! \mathbf{W}_{\text{\tiny{WPE}}}(f)^H \tilde{\mathbf{x}}(t\!\!-\!\!D, f) \right\|_2^2} {\lambda(t, f)}}. \label{equation:wpe_constraint}
\end{equation}
leads to alternating updates between the WPE filter parameters,
\begin{align}  \label{equation:dnn_wpe_filter}
 \mathbf{W}_{\text{\tiny{WPE}}}(f) = {\left(\sum_{t}\frac{\tilde{\mathbf{x}}(t-D, f) \tilde{\mathbf{x}}(t-D, f)^{H}}{\lambda(t,f)}\right)^{-1}} \nonumber \\ \left(\sum_{t}\frac{\tilde{\mathbf{x}}(t-D, f) \mathbf{x}(t, f)^{H}}{\lambda(t,f)}\right)
\end{align}
and the residual signal power given the current WPE filter
\begin{equation} \label{equation:dnn_wpe_power_average}
    \lambda(t,f) = \frac{1}{R} \left\| \hat{\mathbf{d}}(t, f) \right\|_2^2,
\end{equation}
 where $\|\cdot\|_2$ denotes the Euclidean norm. The above alternating estimation procedure iterates until convergence. 
 
 Recent deep neural network extension to WPE led to the DNN-WPE approach \cite{kinoshita2017neural}, where the filtered signal power $\lambda(t,f)$ is estimated using DNN (e.g. LSTM \cite {kinoshita2017neural}) predicted TF complex mask\footnote{Alternatively using channel dependent predicted mask $M_{\text{\tiny{WPE}}}^{r}(t,f)$ produced comparable performance in practice while increasing the system training time approximately by a factor of 5, and therefore not considered.}
 $M_{\text{\tiny{WPE}}}(t,f) \in \mathbb{C}$. This is given by 
\begin{equation} \label{equation:dnn_wpe_psd}
 \lambda(t,f) = \frac{1}{R} \left \| M_{\text{\tiny{WPE}}}(t,f) \mathbf{x}(t,f) \right \|_2^2 ,
\end{equation}
An example of DNN-WPE based dereverberation is shown in Fig. \ref{fig:dervb_module} (top right, in light blue). 

\vspace{-0.4cm}
\subsection{SpecM Based Dereverberation} \label{subsection:SpecM}
\vspace{-0.1cm}
In addition to DNN-WPE based dereverberation, SpecM based dereverberation is also leveraged in this work.  A neural network based TF spectral transformation between the input reverberant and desired anechoic speech spectrum is learned as follows: 
\vspace{-0.2cm}
\begin{align}
\hat{\mathbf{d}}(t, f) = W_{\text{\tiny{SpecM}}}(t,f) \mathbf{x}(t,f)=M_{\text{\tiny{SpecM}}}(t,f) \mathbf{x}(t,f),\label{equation:specm_filter_2}
\end{align}
where $W_{\text{\tiny{SpecM}}}(t,f) \in \mathbb{C}$ denotes the SpecM filter and $M_{\text{\tiny{SpecM}}}(t,f)\in \mathbb{C}$ is the estimated complex TF-mask for SpecM based dereverberation.

An example of SpecM based speech dereverberation is shown in Fig. \ref{fig:dervb_module} (bottom right, in light yellow). Compared with DNN-WPE, although the SpecM based dereverberation approach can provide perceptually enhanced sounds, it has been reported that the artifacts resulting from deterministic spectral masking introduced a negative impact on downstream speech recognition system performance \cite{yoshioka2015ntt, yoshioka2018multi, yoshioka2018recognizing}.

\vspace{-0.3cm}
\subsection{Audio-visual Speech Dereverberation} 
\vspace{-0.1cm}
\label{subsection:av_dervb}
The audio and video embeddings previously used in the mask-based MVDR speech separation network of Section \ref{section:sep} and Fig. \ref{fig:mvdr_wpd_module} are concatenated\footnote{Alternative audio-visual modality fusion methods, e.g. using the factorized attention based fusion mechanism of Section \ref{subsection:modality_fusion} for speech separation, led to performance degradation in practice and therefore not considered.} 
before being fed into an AV Fusion Block consisting of three TCN modules to produce the integrated audio-visual embeddings (Fig. \ref{fig:dervb_module}, left).

These audio-visual embeddings are then forwarded into linear layers (Fig. \ref{fig:dervb_module}, right, yellow blocks) to estimate the complex TF masks of the desired speech for either DNN-WPE (Fig. \ref{fig:dervb_module}, top right, light blue) or SpecM (Fig. \ref{fig:dervb_module}, bottom right, light yellow) based dereverberation filter estimation. In this work, the dereverberation network is trained in both cases using the MSE loss computed between the filtered and ground-truth anechoic speech spectrum\cite{kinoshita2017neural, tan2020audio, li2022audio}.

\vspace{-0.2cm}
\section{Audio-Visual Separation and Dereverberation} 
\label{section:AV_Sep&Dervb}
In this section, three integrated audio-visual speech separation and dereverberation architectures are proposed.
These include: a) a serial pipelined connection of the audio-visual speech separation component with the following dereverberation module; or b) conversely audio-visual speech dereverberation followed by separation; and c) joint speech separation \& dereverberation using audio-visual mask-based WPD. 

\vspace{-0.2cm}
\subsection{Audio-visual Speech Separation-Dereverberation} \label{subsection:Sep-Dervb}
In the audio-visual speech separation-dereverberation architecture, the multi-channel mixture speech spectra $\mathbf{y}(t,f) \in \mathbb{C}^{R}$ as well as the extracted visual features and the camera captured target speaker's DOA from the Visual Front-end module (e.g. Fig. \ref{fig:mvdr_wpd_module}, bottom left corner, in light green) are first fed into the MVDR separation module as shown in Fig. \ref{fig:mvdr_wpd_module}(a) to produce single-channel outputs, $\hat{S}_{\text{\tiny{MVDR}}}(t, f)$, before being connected to the dereverberation module based on DNN-WPE or SpecM as shown in Fig. \ref{fig:dervb_module} to obtain the final enhanced speech $\hat{d}_{\text{\tiny{MVDR-WPE}}}(t, f) \in \mathbb{C}$ or $\hat{d}_{\text{\tiny{MVDR-SpecM}}}(t, f) \in \mathbb{C}$, respectively.

When DNN-WPE based dereverberation is used, this is computed in a two stage,  pipelined manner as
\begin{align}
& \hat{S}_{\text{\tiny{MVDR}}}(t, f) =\mathbf{w}_{\text{\tiny{MVDR}}}(f)^H \mathbf{y}(t, f), \\
& \hat{d}_{\text{\tiny{MVDR-WPE}}}(t, f) \! = \! \hat{S}_{\text{\tiny{MVDR}}}(t, f) \!\! -  \!\! \mathbf{W}_{\text{\tiny{WPE}}}(f)^{H} \hat{\mathbf{s}}_{\text{\tiny{MVDR}}}(t\!-\!D, f),
\end{align}
where
\vspace{-0.3cm}
\begin{align}
\hat{\mathbf{s}}_{\text{\tiny{MVDR}}}(t\!-\!D, f) \!\!=\!\!\left[\hat{S}_{\text{\tiny{MVDR}}}(t\!-\!D, f), \ldots, \hat{S}_{\text{\tiny{MVDR}}}(t\!-\!D\!-\!L\!+\!1, f)\right]^{T} \nonumber
\end{align}
denotes the enhanced single-channel output of the MVDR beamformer from the past $L$ frames and $\hat{\mathbf{s}}_{\text{\tiny{MVDR}}}(t\!-\!D, f) \in \mathbb{C}^{L}$. Here, $\mathbf{W}_{\text{\tiny{WPE}}}(f) \in \mathbb{C}^{L}$ represents the single-channel WPE filter.
$L$ is the number of filter taps and $D$ denotes the prediction delay parameter in WPE. 

When SpecM based dereverberation is used, the final enhanced single-channel speech spectrum is computed as 
\vspace{-0.1cm}
\begin{align}
& \hat{S}_{\text{\tiny{MVDR}}}(t, f) =\mathbf{w}_{\text{\tiny{MVDR}}}(f)^H \mathbf{y}(t, f), \\
& \hat{d}_{\text{\tiny{MVDR-SpecM}}}(t, f) = W_{\text{\tiny{SpecM}}}(t,f) \hat{S}_{\text{\tiny{MVDR}}}(t, f).
\end{align}

\vspace{-0.8cm}
\subsection{Audio-visual Speech Dereverberation-Separation} \label{subsection:Dervb-Sep}
In contrast to the above, connecting the speech dereverberation and separation modules in a reverse order leads to the audio-visual speech dereverberation-separation architecture.
The sequence of filtering operations of this architecture is performed
as follows:

When using DNN-WPE based dereverberation, the dereverberated multi-channel output $\hat{\mathbf{d}}_{\text{\tiny{WPE}}}(t, f)$ is first produced, before being fed into the MVDR separation filter to produce the final single-channel speech spectrum $\hat{S}_{\text{\tiny{WPE-MVDR}}}(t, f)$ as
\begin{align}
& \hat{\mathbf{d}}_{\text{\tiny{WPE}}}(t, f) =\mathbf{y}(t, f)-\mathbf{W}_{\text{\tiny{WPE}}}(f)^H \tilde{\mathbf{y}}(t-D, f), \\
& \hat{S}_{\text{\tiny{WPE-MVDR}}}(t, f) = \mathbf{w}_{\text{\tiny{MVDR}}}(f)^H  \hat{\mathbf{d}}_{\text{\tiny{WPE}}}(t, f),
\end{align}
where
$\tilde{\mathbf{y}}(t\!-\!D, f) \!\!=\!\! \left[\mathbf{y}(t\!-\!D, f)^{T}, \ldots, \mathbf{y}(t\!-\!D\!-\!L\!+\!1, f)^{T} \right]^{T} \in \mathbb{C}^{LR}$
denotes the stacked vector representation of the input multi-channel mixture speech signal. 

When using SpecM based dereverberation, the above can be expressed as
\vspace{-0.2cm}
\begin{align}
& \hat{\mathbf{d}}_{\text{\tiny{SpecM}}}(t, f) = W_{\text{\tiny{SpecM}}}(t,f) \mathbf{y}(t, f), \\
& \hat{S}_{\text{\tiny{SpecM-MVDR}}}(t, f) = \mathbf{w}_{\text{\tiny{MVDR}}}(f)^H  \hat{\mathbf{d}}_{\text{\tiny{SpecM}}}(t, f).
\end{align}
\begin{figure*}[htbp]
    \centering
    \setlength{\abovecaptionskip}{0pt plus 1pt minus 3pt}
    \includegraphics[scale=0.52]{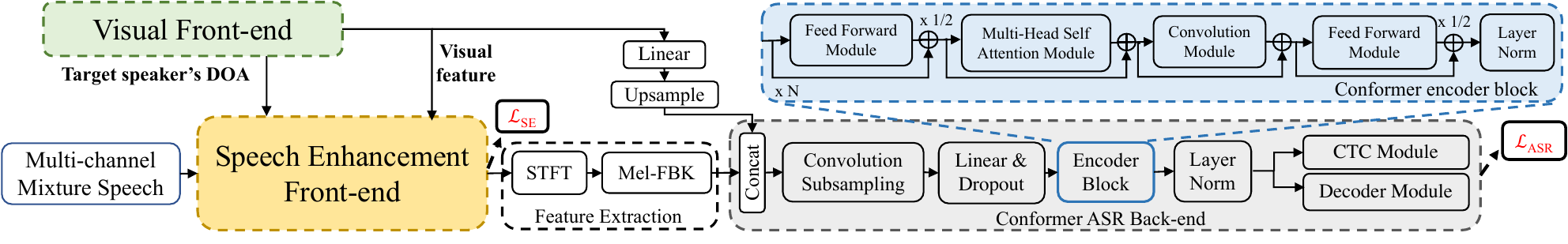}
    \caption{Illustration of an end-to-end audio-visual multi-channel speech separation, dereverberation and recognition system, which integrates the Speech Enhancement Front-end, Visual Front-end, Feature Extraction and Conformer ASR Back-end components.}
    \label{fig:asr_module}
\vspace{-0.5cm}
\end{figure*}
\vspace{-1cm}
\subsection{Audio-visual Joint Speech Separation \& Dereverberation} \label{subsection:Sep&Dervb}
Combining the multi-channel speech separation and dereverberation 
functions into a single convolutional filter leads to a joint 
speech separation and dereverberation architecture, for example, based on
WPD \cite{nakatani2019unified, 2020Jointly, nakatani2019simultaneous} 
and their DNN predicted mask-based variants \cite{zhang2022end}. 

When producing the final enhanced speech spectrum, a single WPD filter  $\tilde{\mathbf{w}}_{\text{\tiny{WPD}}}(f) \in \mathbb{C}^{(L+1)R}$ is applied to the time-delayed multi-channel mixed speech vector stacked by  $\mathbf{y}(t, f) \in \mathbb{C}^{R} $ and $\tilde{\mathbf{y}}(t-D, f)^T  \in \mathbb{C}^{LR} $ as follows:
\begin{equation} \label{equation:wpd_filtering}
\hat{d}(t, f)=\tilde{\mathbf{w}}_{\text{\tiny{WPD}}}(f)^H \left[\mathbf{y}(t, f)^T, \tilde{\mathbf{y}}(t-D, f)^T \right]^T,
\end{equation}

The WPD beamformer is trained to minimize the average weighted power of the filtered signal while satisfying an orthogonal constraint for channel synchronization without distorting the target speech. This is given by 
\begin{align}
& \min _{\tilde{\mathbf{w}}_{\text{\tiny{WPD}}}(f)} \sum_t \frac{\left|\tilde{\mathbf{w}}_{\text{\tiny{WPD}}}(f)^H  \left[\mathbf{y}(t, f)^T, \tilde{\mathbf{y}}(t-D, f)^T \right]^T \right|^2}{\lambda(t, f)}, \\
& \text { subject to }: \tilde{\mathbf{w}}_{\text{\tiny{WPD}}}(f)^H \tilde{\mathbf{g}}(f)=1.
\end{align}
where the signal variance is averaged across $R$ channels as
\begin{equation}
\lambda(t,f) = \frac{1}{R} \sum_{r=1}^R \left | M^{\lambda}_{\text{\tiny{WPD}}}(t,f)Y_r(t,f) \right|^2, \nonumber
\end{equation}
is estimated using DNN predicted TF complex mask of the desired signal $M^{\lambda}_{\text{\tiny{WPD}}}(t, f) \in \mathbb{C}$.
$Y_r(t,f)$ represents the $r$-th component of the multi-channel mixture speech signal $\mathbf{y}(t,f)$.
$\tilde{\mathbf{g}}(f)=\left[\mathbf{g}(f)^T, \mathbf{0}, \ldots, \mathbf{0}\right]^T \in \mathbb{C}^{(L+1)R}$ is the padded steering vector which is composed of a steering vector $\mathbf{g}(f) \in \mathbb{C}^{R}$ and the others $\mathbf{0} \in \mathbb{C}^{R}$ vectors. It can be shown that the solution of the above WPD convolutional beamformer is:
\begin{align}
{\small \tilde{\mathbf{w}}_{\text{\tiny{WPD}}}(f)\!=\!\frac{\boldsymbol{\Phi}_{\tilde{y}}(f)^{-1} \tilde{\mathbf{g}}(f)}{\tilde{\mathbf{g}}(f)^H \boldsymbol{\Phi}_{\tilde{y}}(f)^{-1} \tilde{\mathbf{g}}(f)}\!=\!\frac{\boldsymbol{\Phi}_{\tilde{y}}(f)^{-1} \boldsymbol{\Phi}_{\tilde{\mathbf{x}}}(f)}{\operatorname{tr}\left(\boldsymbol{\Phi}_{\tilde{y}}(f)^{-1} \boldsymbol{\Phi}_{\tilde{\mathbf{x}}}(f)\right)}\tilde{\mathbf{u}}_{r}}, \label{equation:wpd_filter_2}
\end{align}
where the target speaker and power normalized spatial-temporal PSD matrices are
\begin{align}
& {\small \boldsymbol{\Phi}_{\tilde{x}}(f) \!=\!\frac{\sum_{t}\left(M^{\tilde{x}}_{\text{\tiny{WPD}}}(t, f) \widetilde{\mathbf{y}}(t, f)\right)\left(M^{\tilde{x}}_{\text{\tiny{WPD}}}(t, f) \widetilde{\mathbf{y}}(t, f)\right)^{H}}{\sum_{t} M^{\tilde{x}}_{\text{\tiny{WPD}}}(t, f) \left(M^{\tilde{x}}_{\text{\tiny{WPD}}}(t, f)\right)^{*}}}, \label{equation:wpd_speech_PSD} \\
& {\small \boldsymbol{\Phi}_{\tilde{y}}(f)=\sum_{t}\frac{\widetilde{\mathbf{y}}(t, f) \widetilde{\mathbf{y}}(t, f)^H}{\lambda(t, f)}}, \label{equation:wpd_time_varying_power_PSD} 
\end{align}
and $\widetilde{\mathbf{y}}(t, f) = \left[\mathbf{y}(t, f)^T, \tilde{\mathbf{y}}(t-D, f)^T \right]^T \in \mathbb{C}^{(L+1)R}$.
$\tilde{\mathbf{u}}_{\mathbf{r}} = \left [\mathbf{u}_r, \mathbf{0}, \ldots, \mathbf{0} \right]^T$ is the padded reference vector. $M^{\tilde{x}}_{\text{\tiny{WPD}}}(t, f) \in \mathbb{C}$ denotes the complex TF mask of target speech.

An example of mask-based WPD is illustrated in Fig. \ref{fig:mvdr_wpd_module}(b) (bottom right, in light blue).
The same audio-visual embeddings that are used in mask-based MVDR separation module (Fig. 1, top right, light yellow) are now fed into three TCN based Target Speech Block
and Time-varying Power Block for WPD filtering.
Their respective outputs are then fed into the separate linear layers to estimate the complex TF masks $M^{\tilde{x}}_{\text{\tiny{WPD}}}(t, f) \in \mathbb{C}$ and $M^{\lambda}_{\text{\tiny{WPD}}}(t, f) \in \mathbb{C} $ required for the computation of the two spatial-temporal PSD matrices and finally the WPD filter parameters. 
The entire mask-based WPD network is trained using an equally weighted interpolation between the SISNR and MSE losses
to perform joint speech separation \& dereverberation. 

\vspace{-0.2cm}
\section{Audio-visual Multi-Channel Speech Recognition} \label{section:comformer_asr}
In this section, the Conformer-based audio-visual speech recognition back-end and its further integration with the speech enhancement front-end are introduced.
\vspace{-0.4cm}
\subsection{Audio-visual Conformer Speech Recognition Back-end}
\vspace{-0.1cm}
As shown in Fig. \ref{fig:asr_module} (bottom left), the enhanced speech waveform produced by the speech separation and dereverberation front-ends of Sections \ref{section:sep}, \ref{section:dervb} and \ref{section:AV_Sep&Dervb} is fed through a STFT transform before log Mel-filterbank (Mel-FBK) audio features are calculated.
As is also shown in Fig. \ref{fig:asr_module} (top left), the visual features extracted from the Visual Front-end are forwarded into a linear layer before being up-sampled to be time synchronised with the Mel-FBK audio frames. 
Finally, the audio and visual features are concatenated and fed into the ASR back-end.

The Conformer ASR back-end \cite{gulati2020conformer, guo2021recent} comprises a Conformer encoder and a Transformer decoder. 
The Conformer encoder has one convolutional subsampling module, and a linear layer with dropout operation followed by stacked encoder blocks. 
The internal components of each Conformer encoder block include: a position-wise feed-forward network module, a multi-head self-attention module, a convolution module, and a final position-wise feed-forward network module at the end. 
All the encoder blocks additionally undergo layer normalization and residual connections. 
Fig. \ref{fig:asr_module} (right) shows an example of a Conformer ASR system, where the backbone model architecture is in the grey colored part (Fig. \ref{fig:asr_module}, bottom right). The detailed encoder block compositions are in the blue colored part (Fig. \ref{fig:asr_module}, top right). 
The following multi-task criterion interpolation between the CTC and attention error costs \cite{watanabe2017hybrid} is utilized in Conformer model training,
\vspace{-0.2cm}
\begin{equation} \label{equation:interpolation}
\mathcal{L_\text{ASR}}=(1-\beta) \mathcal{L}_{a t t} + \beta \mathcal{L}_{c t c},
\end{equation}
where $\beta \in[0,1]$ is a tunable hyper-parameter and empirically set as $0.3$ for training and $0.4$ for recognition in this paper.

\vspace{-0.4cm}
\subsection{Integration of Speech Enhancement and Recognition} \label{subsection:joint_fine-tuning}
\vspace{-0.1cm}
Traditionally, the speech enhancement front-end and recognition back-end components are optimized separately and used in a pipelined manner \cite{yoshioka2018recognizing, ochiai2020beam, drude2018integrating, erdogan2016improved, yoshioka2018multi}. 
However, two issues arise with this pipelined approach: 
\textbf{1)} the learning cost function mismatch between speech enhancement front-end and recognition back-end components is not addressed; 
\textbf{2)} the artifacts brought by the speech enhancement front-end can lead to ASR performance degradation. 
To this end, a tight integration of the audio-visual speech separation, dereverberation and recognition components via joint fine-tuning \cite{von2020multi, subramanian2020far, xu2019joint, shao2022multi, heymann2019joint, kumar2022end, zhang2022end, yu2021audio, wang2022sjtu} is considered in this paper.
Three fine-tuning methods are investigated: 
\textbf{a)} only fine-tuning the back-end ASR component using the enhanced speech outputs while the front-end remains unchanged; 
\textbf{b)} end-to-end jointly fine-tuning the entire system including the speech enhancement front-end and the recognition back-end components using the ASR cost function; 
\textbf{c)} end-to-end jointly fine-tuning the entire system using a multi-task criterion interpolation between the speech enhancement and recognition cost functions as follows:
\vspace{-0.2cm}
\begin{equation} \label{equation:interpolate_weight}
\mathcal{L}=(1 - \gamma) \mathcal{L}_{\text{ASR}} + \gamma \mathcal{L}_{\text{SE}},
\end{equation}
where $\gamma$ is empirically set as 0.5 in the experiments unless otherwise stated.
The precise form of the speech enhancement loss function, $\mathcal{L}_{\text{SE}}$, is determined by the underlying integrated front-end architectures being used, as described in Section \ref{section:AV_Sep&Dervb}. This is expressed as follows: 
\textbf{a)} $\mathcal{L}_{\text{SE}}\!=\!\mathcal{L}_{\text{SISNR}}$ for audio-visual speech separation followed by dereverberation, as in Section \ref{subsection:Sep-Dervb}; 
\textbf{b)} $\mathcal{L}_{\text{SE}}\!=\!\mathcal{L}_{\text{MSE}}$ for audio-visual speech dereverberation followed by separation, as in Section \ref{subsection:Dervb-Sep}; 
and \textbf{c)} $\mathcal{L}_{\text{SE}}\!=\!\mathcal{L_\text{SISNR}} + \mathcal{L_\text{MSE}}$ for joint speech separation \& dereverberation in Section \ref{subsection:Sep&Dervb}. 

\vspace{-0.2cm}
\section{Experimental Setup}
This section is organized as follows. 
Section \ref{subsection:lrs2_corpus} gives the details of the LRS2 corpus. 
The simulated and replayed multi-channel mixture speech datasets are described in Section Section \ref{subsection:simulated_mixture_speech} and  \ref{subsection:replayed_mixture_speech}, respectively. 
Section \ref{subsection:baseline_system} presents the performance of the baseline single-channel ASR and AVSR systems on mixture speech. 
Finally, two important implementation issues that affect the performance of the proposed audio-visual multi-channel speech separation, dereverberation and recognition systems are discussed in Section \ref{subsection:implementation_detail}. 

\vspace{-0.4cm}
\subsection{LRS2 Corpus} \label{subsection:lrs2_corpus}
\vspace{-0.1cm}
The Oxford LRS2 corpus \cite{son2017lip} is 
one of the largest publicly available corpora for audio-visual speech recognition.
This corpus consists of news and talk shows from BBC programs. 
This is a challenging AVSR task since it contains thousands of speakers with large variations in head pose. 
The LRS2 corpus is divided into four subsets, i.e. Pre-train, Train, Validation and Test sets. In our experiments, the official Pre-train and Train data sets are combined for model training.

\vspace{-0.4cm}
\subsection{Simulated Overlapped and Reverberant Speech} \label{subsection:simulated_mixture_speech}
Since there is no publicly available audio-visual multi-channel mixture speech corpus, we simulated the multi-channel mixture speech with overlapping and reverberation based on the LRS2 corpus in the experiments. 
Details of the simulation process are described in Algorithm 1.
A 15-channel symmetric linear array with non-even inter-channel spacing [7,6,5,4,3,2,1,1,2,3,4,5,6,7]cm is used in the simulation process.
843 point-source noises \cite{ko2017study} and 20000 room impulse responses (RIRs) generated by the image method \cite{habets2006room} in 400 different simulated rooms are used in our experiment. 
The distance between a sound source and the microphone array center is uniformly sampled from a range of 1m to 5m and the room size ranges from 4m$\times$4m$\times$3m to 10m$\times$10m $\times$6m (length$\times$width$\times$height).
The reverberation time $T_{60}$ is uniformly sampled from a range of 0.14s to 0.92s. 
The average overlapping ratio is around 80\%.
The signal-to-noise ratio (SNR) is uniformly sampled from \{0, 5, 10, 15, 20\}dB, and the signal-to-interference ratio (SIR) is uniformly sampled from \{-6, 0, 6\}dB. 
In addition, the angle difference relative to the microphone array between the target and interfering speakers is uniformly sampled from four ranges of the angle difference
\{[$0^{\circ}$, $15^{\circ}$), [$15^{\circ}$, $45^{\circ}$), [$45^{\circ}$, $90^{\circ}$), [$90^{\circ}$, $180^{\circ}$)\}. 
The final simulated multi-channel datasets contain three subsets with 96997, 4272 and 4972 utterances respectively for training (91.37 hours), validation (2.59 hours) and test (2.32 hours).

\begin{algorithm}[t]
    \caption{Multi-channel mixture speech simulation} 
	\label{alg:algorithm1}
      {
	\KwIn{single-channel anechoic LRS2 corpus}
	\KwOut{multi-channel mixture speech}  
        \ForEach{utterance \textnormal{in LRS2}}{
            1) Uniformly sample an interfering utterance from another speaker in the LRS2 corpus;\\
            2) Uniformly sample a room size from 4m$\times$4m$\times$3m to 10m$\times$10m$\times$6m;\\
            3) Uniformly sample a $T_{60}$ from 0.14s to 0.92s;\\
            4) Uniformly sample a microphone array position in the room;\\
            5) Uniformly sample two speakers' positions while the distance between each speaker and the array is within the range of 1m to 5m;\\
            6) Uniformly sample an angle difference from \{[$0^{\circ}$, $15^{\circ}$), [$15^{\circ}$, $45^{\circ}$), [$45^{\circ}$, $90^{\circ}$), [$90^{\circ}$, $180^{\circ}$) \}; \\
            \While{\textnormal{the angle difference of the target and interfering speakers relative to the microphone array not in the selected range}}{
                7) Re-sample the interfering speaker's position;\\
            }
            8) Generate two multi-channel RIRs for the target and interfering speakers using the above settings and applying the image method\cite{habets2006room};\\
            9) Convolve each single-channel anechoic speech of current utterance with the corresponding multi-channel RIRs to simulate room reverberation; \\
            10) Uniformly sample a SIR from \{-6, 0, 6\} dB; \\
            11) Scale the target and interfering sources with the sampled SIR;\\
            12) Uniformly sample a noise from a total of 843 point-source noise types\cite{ko2017study};\\
            13) Add two scaled speaker speech signals along with the selected noise under \{0, 5, 10, 15, 20\}dB SNR to obtain the final multi-channel mixture (overlapped, noisy and reverberant) speech.
            }
            }
\end{algorithm}

\vspace{-0.4cm}
\subsection{Replayed Mixture Speech}\label{subsection:replayed_mixture_speech}
\vspace{-0.1cm}
To further evaluate the performance of the proposed approach in a more realistic application environment, a replayed test set \cite{yu2021audio} with 1200 utterances (0.5 hours) of LRS2 Test set recorded in a 10m$\times$5m$\times$3m meeting room is also used in our experiments.
Two loudspeakers are used to replay different utterances simultaneously to produce mixture speech. 
The geometric specification  of the microphone array used during recording is the same as that used in the simulation.
The target and interfering speakers are located at the following directions relative to the microphone array, i.e. \{$15^{\circ}$/$30^{\circ}$, $45^{\circ}$/$30^{\circ}$, $75^{\circ}$/$30^{\circ}$, $105^{\circ}$/$30^{\circ}$, $30^{\circ}$/$60^{\circ}$, $90^{\circ}$/$60^{\circ}$, $120^{\circ}$/$60^{\circ}$, $150^{\circ}$/$60^{\circ}$\}, where the distance between the loudspeakers and microphones ranges from 1m to 1.5m. 
In the replayed data, the target speaker’s DOA is captured by a $180^{\circ}$ camera \cite{yu2021audio}. 
The average overlapping ratio of the replayed mixture speech is around 80\% and SIR is around 1.5dB. 

\vspace{-0.4cm}
\subsection{Baseline System Description}\label{subsection:baseline_system}
\vspace{-0.1cm}
\textbf{1) Speech Enhancement Front-end:}
The 257-dimensional complex spectrum of each channel is extracted using a 512-point STFT with a 32ms square-root Hanning window and 16ms frame rate (e.g. Fig. \ref{fig:mvdr_wpd_module}, top left corner).
The AF and IPD features are computed using 9 microphone pairs \{1/15, 2/14, 3/13, 1/7, 12/4, 11/5, 12/8, 7/10, 8/9\} to sample different spacing between microphones following \cite{yu2021audio}.
For each Dilated 1D Conv Block in a TCN module (Fig. \ref{fig:tcn_visual_block_module}, left), the number of channels in the 1$\times$1 Conv layer is set to 256.
The kernel size of the D-Conv layer is set to 3, with 512 channels.
The output dimension of the linear layer is set to 257.

\textbf{2) Visual Front-end:}
The original 160$\times$160 dimensional video frames in the LRS2 datasets are centrally cropped by a 112$\times$112 dimensional window and then up-sampled to be time synchronised with the audio frames via linear interpolation. 
The Visual Front-end (e.g. Fig. \ref{fig:mvdr_wpd_module}, bottom left corner, in light green) uses the same hyper-parameter settings as described in \cite{afouras2018deeplip}. 
In addition, the number of the acoustic subspaces $K$ is set to 10 with $\mathbf{P}_k^a \in \mathbb{R}^{256 \times 256}$ and $\mathbf{P}^v \in \mathbb{R}^{10 \times 256}$ in the factorized attention layer \cite{gu2020multi}. 

\textbf{3) Recognition Back-end:}
The 80-dimensional log Mel-FBK features extracted using a 25ms window and 10ms frame rate serve as the inputs to the recognition back-end. 
The baseline Conformer models consist of 12 encoder and 6 decoder blocks following the ESPnet recipe\footnote{github.com/espnet/espnet/blob/master/egs/lrs2/asr1/run.sh}. 
Each encoder or decoder block is configured with 4-head attention of 256 dimensions and 2048 feed-forward hidden units. 
The convolutional sub-sampling module includes two 2D
convolutional layers with a stride of 2, each followed by a ReLU activation. 
500 byte-pair-encoding (BPE) tokens are used as decoder outputs. 
All models are trained using NVIDIA A40 GPU cards\footnote{The jointly fine-tuned speech enhancement front-end and recognition back-end systems in Table V are trained using one thread on a single Nvidia A40 GPU with a batch size of 24 and the GPU memory usage vary from 32G to 43G maximum.}.

\begin{table}[!t]
\centering
\setlength{\abovecaptionskip}{0pt plus 1pt minus 4pt}
\caption{
Performance of single-channel ASR and AVSR systems (without speech enhancement front-end) trained and evaluated on anechoic, reverberant-only and mixture speech. "Simu" and "Replay" denote the simulated and replayed evaluation datasets of Section \ref{subsection:simulated_mixture_speech} and Section \ref{subsection:replayed_mixture_speech}.\label{tab:table1}}
\resizebox{0.9\columnwidth}{!}{
\begin{tabular}{c|c|c|c|c} 
\toprule[1.5pt]
\multirow{2}{*}{Sys.} & \multirow{2}{*}{Data} & \multirowcell{2}{+Visual \\ Features} & \multicolumn{2}{c}{WER(\%)} \\
\cline{4-5} 
 & & & Simu &  \footnotesize Replay \\
\midrule[1pt]
1 & \multirow{2}{*}{Anechoic} & \xmark & 8.8  & \multirow{2}{*}{-} \\ 
2& &\cmark & 7.3 &   \\
\midrule[1pt]
3 & \multirow{2}{*}{Reverberant-only} & \xmark & 13.8  & \multirow{2}{*}{-} \\ 
4 & & \cmark & 10.5 &  \\
\midrule[1pt]
5 & \multirow{2}{*}{Mixture of raw channel 1 } & \xmark & 57.5  & 58.6 \\ 
6 & & \cmark & 25.2  & 22.6\\
\bottomrule[1.5pt]
\end{tabular}}
\vspace{-0.6cm} 
\end{table}

\textbf{4) Performance of Speech Recognition without Speech Enhancement Front-end:} 
Table \ref{tab:table1} presents the WER results of the single-channel input based Conformer ASR and AVSR systems (without using a microphone array and any speech enhancement front-end) on the anechoic, reverberant-only and mixture speech.
It can be observed that using visual information can consistently improve the recognition performance over the audio-only ASR systems by up to \textbf{1.5\% absolute} (\textbf{17.0\% relative}) WER reduction on the anechoic speech (sys. 2 vs. sys. 1) and \textbf{3.3\% absolute} (\textbf{23.9\% relative}) WER reduction on the reverberant-only speech (sys. 4 vs. sys. 3).
In particular, the AVSR system significantly outperforms the audio-only ASR system (sys. 6 vs. sys. 5) by up to \textbf{32.3\%} and \textbf{36.0\% absolute} (\textbf{56.2\%} and \textbf{61.4\% relative}) WER reductions on the simulated and replayed mixture speech respectively. 

\vspace{-0.4cm}
\subsection{Implementation Details} \label{subsection:implementation_detail}
\vspace{-0.1cm}
\begin{table}[!t]
\centering
\setlength{\abovecaptionskip}{0pt plus 1pt minus 4pt}
\caption{Performance of three integrated speech enhancement front-end architectures with different numbers of filter taps ($L$) on simulated mixture speech for single-channel DNN-WPE, multi-channel DNN-WPE and mask-based WPD modules used in audio-only speech enhancement front-ends. \label{tab:table2}}
\resizebox{1.0\columnwidth}{!}{
\begin{tabular}{c|c|c|c|c} 
\toprule[1.5pt]
\multirow{3}{*}{Sys.} & \multirowcell{3}{\scriptsize{Filter taps}\\(\scriptsize $L$)}& \multicolumn{3}{c}{PESQ$(\uparrow)$ / STOI$(\uparrow)$ / SRMR$(\uparrow)$} \\
\cline{3-5} 
&  & \multirowcell{2}{Sep. $\rightarrow$ Dervb. \\ \scriptsize{(Single-channel DNN-WPE)}}  & \multirowcell{2}{Dervb. $\rightarrow$ Sep. \\ \scriptsize{(Multi-channel DNN-WPE)}} & \multirowcell{2}{Joint  Sep. \& Dervb. \\ \scriptsize{(Mask-based WPD)}}   \\
 &&&& \\
\midrule[1pt]
1& 1 & 2.21/72.07/5.32 & 2.44/79.63/6.31 & \textbf{2.42/76.63/6.64} \\ 
2& 2 & 2.22/72.42/5.29 & \textbf{2.46/79.75/6.44} & 2.40/76.64/6.83 \\ 
3& 3 & 2.23/72.69/5.32 & 2.45/79.66/6.50 & 2.40/76.51/6.97 \\
4& 4 & 2.23/72.86/5.35 & 2.45/79.53/6.57  & 2.36/76.10/7.04 \\ 
5& 5 & 2.24/72.98/5.39 & 2.44/79.32/6.60 & 2.34/75.78/7.08 \\ 
6& 7 & 2.24/73.20/5.45 & 2.41/78.47/6.72 & 2.30/75.05/7.11 \\ 
7& 9 & 2.24/73.35/5.51 & 2.38/77.87/6.70 & 2.27/74.48/7.16 \\ 
8& 12 & 2.25/73.53/5.58 & 2.34/76.73/6.80 &2.20/73.28/7.12  \\
9 & 15 & 2.25/73.65/5.64 & 2.28/75.20/6.83 & 2.12/71.74/6.90 \\ 
10& 18 & \textbf{2.25/73.73/5.70} & 2.24/74.18/6.90  &2.06/70.39/6.66\\
11& 21 & 2.25/73.71/5.75 &  2.18/72.67/6.84 & 1.98/68.90/6.48 \\ 
12& 24 & 2.25/73.71/5.79 &  2.11/71.09/6.90  & 1.87/66.20/6.02\\ 
13& 27 & 2.25/73.70/5.82 &  2.02/68.96/6.72 & 1.81/64.60/5.83 \\ 
\bottomrule[1.5pt]
\end{tabular}}
\vspace{-0.6cm}
\end{table}
\textbf{1) Number of Filter Taps:}
The number of filter taps $L$ used in WPE and WPD approaches has a huge impact on the quality of the enhanced speech and the downstream recognition performance. 
A set of ablation studies on the settings of filter taps $L$ are conducted for each of the three integrated speech separation and dereverberation front-end architectures of Section \ref{section:AV_Sep&Dervb} (i.e. ``Sep. $\rightarrow$ Dervb", ``Dervb. $\rightarrow$ Sep." and ``Joint Sep. \& Dervb." denote the speech separation followed by dereverberation, speech dereverberation followed by separation and joint speech separation \& dereverberation, respectively.) 
These are shown in Table \ref{tab:table2} for audio-only speech enhancement.
Considering the speech enhancement performance in terms of PESQ, STOI and SRMR scores, the number of filter taps for single-channel DNN-WPE, multi-channel DNN-WPE and mask-based WPD are respectively chosen and fixed as 18 (sys. 10), 2 (sys. 2) and 1 (sys. 1) in the following experiments. 
In addition, the prediction delay $D$ is empirically set to 2 for DNN-WPE and mask-based WPD.

\begin{table}[!t]
\centering
\setlength{\abovecaptionskip}{0pt plus 1pt minus 4pt}
\caption{
Performance of speech enhancement front-ends with different diagonal variance flooring ($\varepsilon$) on simulated mixture speech for mask-based MVDR, single-channel DNN-WPE, multi-channel DNN-WPE and mask-based WPD used in audio-only speech enhancement front-ends. \label{tab:table3}}
\resizebox{1.0\columnwidth}{!}{
\begin{tabular}{c|c|c|c|c|c} 
\toprule[1.5pt]
\multirow{4}{*}{Sys.} & \multirowcell{4}{Variance \\ flooring \\ ($\varepsilon$)} &\multicolumn{4}{c}{PESQ$(\uparrow)$ / STOI$(\uparrow)$ / SRMR$(\uparrow)$ } \\
\cline{3-6} 
 &  & \multirowcell{3}{Sep.\\ \scriptsize{(Mask-based} \\ \scriptsize{MVDR)}}& \multirowcell{3}{Sep. $\rightarrow$ Dervb. \\ \scriptsize{(Single-channel} \\\scriptsize{DNN-WPE)}}  & \multirowcell{3}{Dervb. $\rightarrow$ Sep. \\ \scriptsize{(Multi-channel} \\ \scriptsize{DNN-WPE)}} & \multirowcell{3}{Joint  Sep. \& Dervb. \\ \scriptsize{(Mask-based WPD)}}   \\
 &&&&& \\
 &&&&& \\
\midrule[1pt]
1 &$10^{-1}$& 1.89/63.41/4.36 &  2.21/71.98/5.67 & 2.36/77.68/6.01  & 2.11/67.85/5.47\\ 
2 &$10^{-3}$& 2.08/68.50/4.77  &  2.24/73.39/5.88 & 2.44/79.37/6.24 & 2.36/75.38/6.45 \\
3 &$10^{-4}$& 2.17/70.39/5.34   &  2.25/73.64/5.80 & 2.43/79.25/6.18 & \textbf{2.42/76.63/6.64} \\
4 &$10^{-5}$& \textbf{2.21/71.30/5.45}  &  \textbf{2.25/73.73/5.70} & 2.45/79.68/6.40 & 1.96/61.45/6.29\\ 
5 &$10^{-6}$& 2.19/71.24/5.46   &  2.25/73.74/5.65 & \textbf{2.46/79.75/6.44} & 1.55/45.13/4.84 \\
6 &$10^{-7}$& 2.16/70.92/5.39  &  2.25/73.75/5.64 & 2.44/79.62/6.56 & 1.63/48.27/4.74\\ 
7 &$10^{-9}$& 1.99/67.02/5.23  &  2.25/73.74/5.64 & 2.25/73.67/5.95 & 1.51/43.99/4.27 \\ 
\bottomrule[1.5pt]
\end{tabular}}
\vspace{-0.6cm}
\end{table}
\textbf{2) Matrix Inversion:}
The inversion of the PSD matrices for MVDR and WPD (Eqn. (\ref{equation:mvdr_filter_2}) and Eqn. (\ref{equation:wpd_filter_2})) and the temporal correlation matrix for WPE (Eqn. (\ref{equation:dnn_wpe_filter})) are prone to numerical issues when they are ill-conditioned or singular. 
To this end, the diagonal variance flooring approach \cite{zhang2022end} is utilized in this work.
A complex PSD or correlation matrix $\boldsymbol{\Phi}$ is floored as $\boldsymbol{\Phi}^{\prime}=\boldsymbol{\Phi}+\varepsilon \operatorname{tr}(\boldsymbol{\Phi}) \mathbf{I}$ before inversion, where a flooring scaling term $\varepsilon$ needs to be set, and $\mathbf{I}$ is the identity matrix. 
In addition, a more stable complex matrix inversion algorithm \cite{smith1974note} is adopted in this paper.  
A set of ablation studies on the setting of the flooring scaling $\varepsilon$ is shown in Table \ref{tab:table3} for audio-only speech enhancement front-end systems with different separation only or integrated (separation and dereverberation) architectures. 
Based on the PESQ, STOI and SRMR scores, $10^{-5}$ (sys. 4), $10^{-5}$ (sys. 4), $10^{-6}$ (sys. 5) and $10^{-4}$ (sys. 3) are selected as the optimal values of the diagonal variance flooring scaling $\varepsilon$ for mask-based MVDR, single-channel DNN-WPE, multi-channel DNN-WPE and mask-based WPD respectively in the following experiments.

\vspace{-0.2cm}
\section{Experimental Results}
\begin{table*}[!t]
\vspace{-0.2cm}
\centering
\setlength{\abovecaptionskip}{0pt plus 1pt minus 4pt}
\caption{
Performance of integrated architectures for audio-visual multi-channel speech separation (``Sep."), dereverberation (``Dervb.") and recognition (``RECG.") on the LRS2 simulated multi-channel mixture dataset.
``Arch.", ``AF", ``SpecM", ``Conf." and ``Avg."  denote the architecture, angle feature, spectral mapping, Conformer and average, respectively.
[$a^{\circ}$, $b^{\circ}$) denotes the range of inter-speaker angle difference between the target and interfering speakers relative to the microphone array.
``$\ast$" and ``$\dagger$" represent a statistically significant WER difference over the corresponding audio-only baseline systems (sys. 5,12,19,26,33) and audio-only dereverberation baseline systems (sys. (5,7,9),(12,14,16),(19,21,23),(26,28,30)), respectively.
}
\label{tab:table4}
\resizebox{2.0\columnwidth}{!}{
\begin{tabular}{c|c|c|c|cc|c|c|c|c|c|c} 
 \toprule[2pt]
 \multirow{3}{*}{\scriptsize Arch.} & \multirow{3}{*}{\scriptsize Sys.} & \multirow{3}{*}{\scriptsize +AF} &  \multicolumn{4}{c|}{\scriptsize +Visual Features} & \multicolumn{5}{c}{PESQ$(\uparrow)$ / STOI$(\uparrow)$ / SRMR$(\uparrow)$ / WER$(\downarrow)$}  \\
 \cline{4-12} 
 &  &  & \tiny Sep. &  \multicolumn{2}{c|}{\tiny Dervb.}  & \tiny Recg. & \multirow{2}{*}{\scriptsize [$0^{\circ}$, $15^{\circ}$)} & \multirow{2}{*}{\scriptsize  [$15^{\circ}$, $45^{\circ}$)}\ & \multirow{2}{*}{\scriptsize [$45^{\circ}$, $90^{\circ}$)}  & \multirow{2}{*}{\scriptsize  [$90^{\circ}$, $180^{\circ}$)}  &\multirow{2}{*}{\scriptsize Avg.}\\
& & & \tiny(MVDR) &\tiny (DNN-WPE) & \tiny (SpecM) & \tiny(Conf.) & & & & & \\
 \midrule[1.5pt]
 \multicolumn{6}{c|}{\multirow{2}{*}{Mixture of raw channel 1}} &\xmark  & \scriptsize{1.54/53.98/3.58/57.9}& \scriptsize{1.53/53.48/3.57/57.4}  & \scriptsize{1.53/53.58/3.58/57.8} & \scriptsize{1.54/54.08/3.60/57.0} & \scriptsize{1.54/53.78/3.58/57.5} \\
  \multicolumn{6}{c|}{}&\cmark  & \scriptsize{1.54/53.98/3.58/25.9}& \scriptsize{1.53/53.48/3.57/24.7}  & \scriptsize{1.53/53.58/3.58/25.6} & \scriptsize{1.54/54.08/3.60/24.5} & \scriptsize{1.54/53.78/3.58/25.2} \\
\midrule[1.5pt]
\multirowcell{4}{ \textbf{Sep.} \\ \scriptsize (MVDR only)} & 1 & \cmark & \xmark  & \multicolumn{2}{c|}{-} & \xmark  & \scriptsize{1.87/62.07/5.03/51.4} & \scriptsize{2.23/71.90/5.50/29.3} &  \scriptsize{2.35/74.95/5.58/22.8} & \scriptsize{2.39/76.28/5.67/21.6} & \scriptsize{2.21/71.30/5.45/31.3} \\
  & 2& \xmark & \cmark  & \multicolumn{2}{c|}{-} &  \xmark & \scriptsize{2.20/71.35/5.34/28.6} & \scriptsize{2.29/73.39/5.52/24.9} &  \scriptsize{2.33/74.41/5.52/23.9} & \scriptsize{2.35/75.19/5.60/22.4} & \scriptsize{2.29/73.59/5.50/25.0} \\
 & 3 & \cmark & \cmark  & \multicolumn{2}{c|}{-}& \xmark & \scriptsize{2.15/70.30/5.32/33.1} & \scriptsize{2.30/73.95/5.63/23.0} &  \scriptsize{2.38/75.71/5.65/21.4} & \scriptsize{2.42/76.96/5.74/19.7} & \scriptsize{2.31/74.23/5.59/24.3}\\
 & 4 & \cmark & \cmark  & \multicolumn{2}{c|}{-} & \cmark & \scriptsize{2.15/70.30/5.32/21.7} & \scriptsize{2.30/73.95/5.63/15.9} &  \scriptsize{2.38/75.71/5.65/14.7} & \scriptsize{2.42/76.96/5.74/13.2} & \scriptsize{2.31/74.23/5.59/16.4}\\
 \midrule[1.5pt]
 \multirowcell{7}{ \textbf{Sep. $\rightarrow$ Dervb.} \\ \scriptsize{(MVDR $\rightarrow$ DNN-WPE)}} & 5 & \cmark & \xmark  & \xmark & & \xmark 
 &\scriptsize{1.91/64.21/5.23/50.1} &\scriptsize{2.26/74.36/5.74/27.9} &\scriptsize{2.39/77.51/5.86/21.6} &\scriptsize{2.44/78.82/5.99/20.5} &\scriptsize {2.25/73.73/5.70/30.0} \\
 & 6 & \cmark & \xmark  & \cmark & & \xmark & 
 \scriptsize {1.91/64.51/5.22/49.6}& 
 \scriptsize {2.26/74.56/5.76/27.4}& 
 \scriptsize {2.39/77.69/5.87/21.1}& 
 \scriptsize {2.44/78.96/6.01/20.3}& 
 \scriptsize {2.25/73.93/5.71/29.6$^{\dagger}$}\\
 & 7 & \xmark & \cmark  & \xmark & & \xmark &\scriptsize{\textbf{2.24}/73.72/5.57/28.5} & \scriptsize{2.33/75.84/5.79/25.2} & \scriptsize{2.37/76.95/5.80/22.9}& \scriptsize{2.40/77.65/5.94/22.2} & \scriptsize{2.34/76.04/5.78/24.7} \\
 & 8 & \xmark & \cmark  & \cmark & & \xmark & \scriptsize{\textbf{2.24}/\textbf{73.81}/\textbf{5.60}/28.2}& \scriptsize{2.33/75.90/5.80/24.5} &\scriptsize{2.38/77.03/5.82/23.2} & \scriptsize{2.40/77.74/5.95/22.3} &\scriptsize{2.34/76.12/5.79/24.5} \\
 & 9 & \cmark & \cmark  & \xmark & & \xmark &
 \scriptsize{2.18/72.69/5.54/31.9}& 
 \scriptsize{\textbf{2.35}/76.53/\textbf{5.92}/23.1}& 
 \scriptsize{\textbf{2.43}/78.43/5.94/20.2}& 
 \scriptsize{\textbf{2.47}/79.63/\textbf{6.09}/18.9}&
 \scriptsize{\textbf{2.36}/76.82/5.87/23.5}\\
 & 10 & \cmark & \cmark  & \cmark & & \xmark & \scriptsize{2.18/72.79/5.55/31.7} & \scriptsize{\textbf{2.35}/\textbf{76.62}/\textbf{5.92}/22.7}  &\scriptsize{2.42/\textbf{78.50}/\textbf{5.95}/20.0} & \scriptsize{\textbf{2.47}/\textbf{79.67}/\textbf{6.09}/18.2} & \scriptsize{\textbf{2.36}/\textbf{76.90}/\textbf{5.88}/23.2$^{\dagger}$}\\
 & 11 & \cmark & \cmark  & \cmark & & \cmark & \scriptsize{2.18/72.79/5.55/\textbf{21.1}} & \scriptsize{\textbf{2.35}/\textbf{76.62}/\textbf{5.92}/\textbf{15.2}}  &\scriptsize{2.42/\textbf{78.50}/\textbf{5.95}/\textbf{14.1}} & \scriptsize{\textbf{2.47}/\textbf{79.67}/\textbf{6.09}/\textbf{13.5}} & \scriptsize{\textbf{2.36}/\textbf{76.90}/\textbf{5.88}/\textbf{16.0}$^{\ast}$} \\
 \cline{1-12}
 \multirowcell{7}{ \textbf{Sep. $\rightarrow$ Dervb.} \\ \scriptsize{(MVDR $\rightarrow$ SpecM)}}  & 12 & \cmark & \xmark & & \xmark & \xmark & \scriptsize{1.95/66.13/7.00/52.7} & \scriptsize{2.37/77.53/7.44/30.7} & \scriptsize{2.51/80.81/7.50/23.6} & \scriptsize{2.57/81.91/7.54/22.5} & \scriptsize{2.35/76.60/7.37/32.4}\\
  
 & 13 & \cmark & \xmark & & \cmark & \xmark & \scriptsize{1.98/67.73/7.16/50.6}& \scriptsize{2.41/78.67/7.66/28.3} & \scriptsize{2.55/81.81/7.71/22.8} & \scriptsize{2.60/82.77/7.73/20.4} & \scriptsize{2.39/77.75/7.56/30.5$^{\dagger}$} \\
 
 & 14 & \xmark & \cmark & & \xmark & \xmark &\scriptsize{2.37/78.10/7.50/31.6}
 &\scriptsize{2.47/80.10/7.65/27.3}
 &\scriptsize{2.52/81.12/7.63/24.9}
 &\scriptsize{2.54/81.60/7.66/24.1}
 &\scriptsize{2.48/80.23/7.61/27.0} \\

 & 15 & \xmark & \cmark & & \cmark & \xmark &\scriptsize{\textbf{2.38}/\textbf{78.36}/\textbf{7.62}/29.9} &\scriptsize{2.48/80.28/7.71/25.3}& \scriptsize{2.53/81.25/7.71/23.8}& \scriptsize{2.55/81.69/7.73/22.7} & \scriptsize{2.49/80.39/7.69/25.4$^{\dagger}$} \\
 
 & 16 & \cmark & \cmark & & \xmark & \xmark 
 & \scriptsize{2.31/76.74/7.43/35.0} 
 & \scriptsize{2.50/80.72/7.72/25.8}
 & \scriptsize{\textbf{2.60}/82.61/7.68/22.3}
 & \scriptsize{\textbf{2.64}/\textbf{83.55}/7.73/20.9}
 & \scriptsize{\textbf{2.51}/80.91/7.64/26.0}\\
 
 & 17 & \cmark & \cmark & & \cmark & \xmark 
 & \scriptsize{2.31/76.98/7.58/33.5} 
 & \scriptsize{\textbf{2.51}/\textbf{80.86}/\textbf{7.84}/23.8}
 & \scriptsize{2.59/\textbf{82.71}/\textbf{7.80}/21.7}
 & \scriptsize{\textbf{2.64}/\textbf{83.55}/\textbf{7.85}/19.2}
 & \scriptsize{\textbf{2.51}/\textbf{81.03}/\textbf{7.77}/24.5$^{\dagger}$} \\
 
 & 18 & \cmark & \cmark & & \cmark & \cmark  
 & \scriptsize{2.31/76.98/7.58/\textbf{22.0}} 
 & \scriptsize{\textbf{2.51}/\textbf{80.86}/\textbf{7.84}/\textbf{16.8}}
 & \scriptsize{2.59/\textbf{82.71}/\textbf{7.80}/\textbf{14.5}}
 & \scriptsize{\textbf{2.64}/\textbf{83.55}/\textbf{7.85}/\textbf{14.4}}
 & \scriptsize{\textbf{2.51}/\textbf{81.03}/\textbf{7.77}/\textbf{16.9}$^{\ast}$} \\
\midrule[1.5pt]
  \multirowcell{7}{ \textbf{Dervb. $\rightarrow$ Sep.} \\ \scriptsize{(DNN-WPE $\rightarrow$ MVDR)}} & 19 & \cmark & \xmark  & \xmark & & \xmark  
 & \scriptsize{2.04/69.13/5.86/47.2} 
 & \scriptsize{2.48/80.58/6.46/24.6}
 & \scriptsize{2.63/84.17/6.68/19.2}
 & \scriptsize{2.68/85.11/6.75/19.2}
 & \scriptsize{2.46/79.75/6.44/27.5} \\
 & 20 & \cmark & \xmark  & \cmark & & \xmark  
 & \scriptsize{2.03/69.57/5.85/46.7} 
 & \scriptsize{2.46/80.46/6.39/24.6}
 & \scriptsize{2.62/83.93/6.66/19.2}
 & \scriptsize{2.67/85.14/6.78/19.1}
 & \scriptsize{2.45/79.78/6.42/27.4}  \\
 & 21 & \xmark & \cmark  & \xmark & &\xmark 
 & \scriptsize{2.39/78.99/6.26/27.4} 
 & \scriptsize{2.53/81.61/6.61/22.0}
 & \scriptsize{2.60/83.23/6.69/20.4}
 & \scriptsize{2.60/83.15/6.71/20.9}
 & \scriptsize{2.53/81.75/6.57/22.7} \\
 & 22 & \xmark & \cmark  & \cmark & &\xmark 
 & \scriptsize{\textbf{2.41}/\textbf{79.37}/\textbf{6.33}/25.8} 
 & \scriptsize{2.54/81.90/6.64/21.4}
 & \scriptsize{2.61/83.33/6.67/19.6}
 & \scriptsize{2.61/83.40/6.74/20.4}
 & \scriptsize{2.54/82.00/6.59/21.8$^{\dagger}$}\\
 & 23 & \cmark & \cmark  & \xmark & &\xmark 
 & \scriptsize{2.34/77.63/6.19/30.3} 
 & \scriptsize{\textbf{2.57}/\textbf{82.62}/\textbf{6.65}/20.2}
 & \scriptsize{\textbf{2.68}/84.89/\textbf{6.81}/17.9}
 & \scriptsize{\textbf{2.71}/85.64/\textbf{6.89}/17.1}
 & \scriptsize{\textbf{2.57}/\textbf{82.69}/\textbf{6.64}/21.4}  \\
 & 24 & \cmark & \cmark  & \cmark & &\xmark
 & \scriptsize{2.33/77.39/6.20/30.7} 
 & \scriptsize{2.55/82.40/6.63/20.3}
 & \scriptsize{2.66/\textbf{84.93}/6.74/17.7}
 & \scriptsize{2.70/\textbf{85.71}/6.85/17.2}
 & \scriptsize{2.56/82.61/6.61/21.4} \\
 & 25 & \cmark & \cmark  & \cmark & &\cmark 
 & \scriptsize{2.33/77.39/6.20/\textbf{20.8}} 
 & \scriptsize{2.55/82.40/6.63/\textbf{15.1}}
 & \scriptsize{2.66/\textbf{84.93}/6.74/\textbf{12.4}}
 & \scriptsize{2.70/\textbf{85.71}/6.85/\textbf{12.3}}
 & \scriptsize{2.56/82.61/6.61/\textbf{15.1}$^{\ast}$} \\
 \cline{1-12}
 \multirowcell{7}{ \textbf{Dervb. $\rightarrow$ Sep.} \\ \scriptsize{(SpecM $\rightarrow$ MVDR)}} & 26 & \cmark & \xmark  & &\xmark & \xmark 
 &\scriptsize{1.82/63.34/5.96/57.0}
 &\scriptsize{2.22/73.58/6.44/32.8}  &\scriptsize{2.43/78.33/6.83/25.6}  &\scriptsize{2.49/79.63/6.94/24.2}  &\scriptsize{2.24/73.72/6.54/34.9} \\
 & 27 & \cmark & \xmark  & &\cmark & \xmark &\scriptsize{1.82/62.99/5.82/57.7}
 &\scriptsize{2.24/74.06/6.43/31.9}  &\scriptsize{2.43/78.55/6.82/24.8}  &\scriptsize{2.49/79.73/6.96/23.8}  &\scriptsize{2.24/73.83/6.51/34.6} \\
 & 28 & \xmark & \cmark  & &\xmark &\xmark &\scriptsize{\textbf{2.17}/72.66/6.39/38.8}  & \scriptsize{2.33/76.11/6.73/29.3}  & \scriptsize{2.45/78.55/6.89/24.9} &  \scriptsize{2.45/78.51/6.92/26.4} &\scriptsize{2.35/76.46/6.73/29.8}   \\
 & 29 & \xmark & \cmark  & &\cmark &\xmark &\scriptsize{2.16/\textbf{72.82}/\textbf{6.44}/36.0}
 &\scriptsize{2.31/75.88/6.72/29.3}  &\scriptsize{2.43/78.39/6.90/25.1}  &\scriptsize{2.41/78.16/6.90/25.1}  &\scriptsize{2.33/76.31/6.74/28.9$^{\dagger}$} \\
 & 30 & \cmark & \cmark  & &\xmark &\xmark & \scriptsize{2.12/71.49/6.28/39.7} & \scriptsize{\textbf{2.35}/76.58/\textbf{6.74}/27.1} & \scriptsize{\textbf{2.49}/\textbf{79.62}/\textbf{6.94}/22.4} &  \scriptsize{\textbf{2.53}/\textbf{80.73}/7.03/22.8} &
\scriptsize{\textbf{2.37}/\textbf{77.11}/6.75/28.0}    \\
 & 31 & \cmark & \cmark  & &\cmark &\xmark &\scriptsize{2.11/71.59/6.35/38.2}
 &\scriptsize{2.34/\textbf{76.65}/6.71/27.6}  &\scriptsize{2.47/79.54/6.92/22.4}  &\scriptsize{2.52/80.61/\textbf{7.04}/21.1}  &\scriptsize{2.36/77.10/\textbf{6.76}/27.3$^{\dagger}$}  \\
 & 32 & \cmark & \cmark  & &\cmark &\cmark &\scriptsize{2.11/71.59/6.35/\textbf{24.9}}
 &\scriptsize{2.34/\textbf{76.65}/6.71/\textbf{16.9}}  &\scriptsize{2.47/79.54/6.92/\textbf{15.3}}  &\scriptsize{2.52/80.61/\textbf{7.04}/\textbf{14.8}}  &\scriptsize{2.36/77.10/\textbf{6.76}/\textbf{18.0}$^{\ast}$}   \\
\midrule[1.5pt]
 \multirowcell{4}{\textbf{Joint Sep. \& Dervb.} \\ \scriptsize{(WPD)}} & 33& \cmark & \multicolumn{3}{c|}{\xmark}  & \xmark & \scriptsize{1.99/65.92/6.06/55.0} & \scriptsize{2.41/76.81/6.60/30.1} & \scriptsize{2.60/81.09/6.83/22.6} & \scriptsize{2.67/82.69/7.07/21.8} &
 \scriptsize{2.42/76.63/6.64/32.4}\\
 & 34 & \xmark & \multicolumn{3}{c|}{\cmark} & \xmark & \scriptsize{\textbf{2.29}/\textbf{74.79}/\textbf{6.44}/34.3} & \scriptsize{2.46/78.21/\textbf{6.78}/25.5} & \scriptsize{2.57/80.26/6.91/22.9} & \scriptsize{2.57/80.48/7.01/22.9} &
 \scriptsize{2.47/78.44/6.78/26.4}\\
 
 & 35 & \cmark & \multicolumn{3}{c|}{\cmark} & \xmark & \scriptsize{2.26/73.78/6.36/37.7} & \scriptsize{\textbf{2.50}/\textbf{79.11}/6.75/25.1} & \scriptsize{\textbf{2.64}/\textbf{82.25}/\textbf{6.95}/20.7} & \scriptsize{\textbf{2.70}/\textbf{83.43}/\textbf{7.13}/20.1} &
 \scriptsize{\textbf{2.53}/\textbf{79.64}/\textbf{6.80}/25.9}\\
 & 36 & \cmark& \multicolumn{3}{c|}{\cmark} & \cmark & \scriptsize{2.26/73.78/6.36/\textbf{24.6}} & \scriptsize{\textbf{2.50}/\textbf{79.11}/6.75/\textbf{16.3}} & \scriptsize{\textbf{2.64}/\textbf{82.25}/\textbf{6.95}/\textbf{13.7}} & \scriptsize{\textbf{2.70}/\textbf{83.43}/\textbf{7.13}/\textbf{13.6}} &
 \scriptsize{\textbf{2.53}/\textbf{79.64}/\textbf{6.80}/\textbf{17.0}$^{\ast}$}\\
\bottomrule[2pt]
\end{tabular}}
\vspace{-0.4cm}
\end{table*}

\vspace{-0.1cm}
In this section, the performance of three integrated audio-visual multi-channel speech separation, dereverberation and recognition architectures of Section \ref{section:AV_Sep&Dervb} are evaluated on the LRS2 simulated and replayed mixture speech datasets.
Section \ref{subsection:pipelined_system} analyses the performance improvements by incorporating visual features into different speech enhancement front-end components as well as the recognition back-end.
After end-to-end joint fine-tuning, the performance of tightly integrated audio-visual speech separation, dereverberation and recognition systems are presented in Section \ref{subsection:joint_system}. 

\begin{table*}[!t]
\centering
\setlength{\abovecaptionskip}{0pt plus 1pt minus 4pt}
\caption{Performance of audio-visual and audio-only multi-channel speech recognition systems after end-to-end joint fine-tuning using ASR cost $\mathcal{L}_{\text{ASR}}$ alone (marked with ``({\rm a})"), or its interpolated with enhancement loss $\mathcal{L}_{\text{ASR}}$ + $\mathcal{L}_{\text{SE}}$ (marked with ``({\rm b})"), on the LRS2 simulated (``Simu") and replayed (``Replay") test sets. 
The original system numbering from Table \ref{tab:table4} is used.
``Avg." is in short for ``average" and ``O.V." for ``overall" results on both simulated and replayed test data.
``$\dagger$", ``$\ast$" and ``${\ddag}$" denote a statistically significant WER difference over the systems without joint fine-tuning (marked with ``-"), the corresponding audio-only baseline systems (sys. 5({\rm b}), 12({\rm b}), 19({\rm b}), 26({\rm b}), 33({\rm b})) and separation-only AVSR baseline system (sys. 4({\rm b})), respectively.}
\label{tab:table5}
\resizebox{2.0\columnwidth}{!}{
\begin{tabular}{c|c|c|cc|cc|cc|cc|c|ccc} 
 \toprule[2pt]
 \multirow{3}{*}{Arch.} & \multirowcell{3}{Sys.}&   \multirowcell{3}{ Jointly Fine-tuning \\ Criterion} & \multicolumn{2}{c|}{PESQ$(\uparrow)$ / STOI$(\uparrow)$ / SRMR$(\uparrow)$}  & \multicolumn{10}{c}{ WER$(\downarrow)$} \\
 \cline{4-15} 
 &  &  &  \multicolumn{2}{c|}{ Avg.} & \multicolumn{2}{c|}{\scriptsize [$0^{\circ}$, $15^{\circ}$)} & \multicolumn{2}{c|}{\scriptsize [$15^{\circ}$, $45^{\circ}$)} & \multicolumn{2}{c|}{\scriptsize [$45^{\circ}$, $90^{\circ}$)}  & \multicolumn{1}{c|}{\scriptsize [$90^{\circ}$, $180^{\circ}$)}  &\multicolumn{3}{c}{\scriptsize Avg.}\\
& & & \scriptsize{\quad Simu} & \scriptsize{\quad Replay} & \scriptsize{Simu} & \scriptsize{Replay} & \scriptsize{Simu} &  \scriptsize{Replay} & \scriptsize{Simu} & \scriptsize{Replay} &  \scriptsize{Simu} & \scriptsize{Simu} & \scriptsize{Replay} & \scriptsize{O.V.}\\
 \midrule[1.5pt]
\multirowcell{6}{\textbf{Sep.} \\ \scriptsize (MVDR only)} & \scriptsize{1} & -  & \scriptsize{2.21/71.30/5.45} & \scriptsize{2.32/77.77/4.31} &  \scriptsize{51.4} & \scriptsize{30.6} & \scriptsize{29.3} & \scriptsize{23.6} & \scriptsize{22.8} & \scriptsize{18.5} & \scriptsize{21.6}  & \scriptsize{31.3} & \scriptsize{23.4} & \scriptsize{27.4}\\

& \scriptsize{1(a)} & \scriptsize{$\mathcal{L}_{\text{ASR}}$} & \scriptsize{2.46/77.72/6.27} & \scriptsize{2.55/81.90/5.35} & \scriptsize{41.5} & \scriptsize{33.0} & \scriptsize{21.1} & \scriptsize{20.7} & \scriptsize{17.0} & \scriptsize{18.2} & \scriptsize{16.2} & \scriptsize{24.0} & \scriptsize{22.8} & \scriptsize{23.4} \\

& \scriptsize{1(b)} &  \scriptsize{$\mathcal{L}_{\text{ASR}}$ + $\mathcal{L}_{\text{SE}}$} & \scriptsize{2.32/74.75/5.77} & \scriptsize{2.40/80.11/4.61} & \scriptsize{42.2} & \scriptsize{28.7} & \scriptsize{22.8} & \scriptsize{20.1} & \scriptsize{18.2} & \scriptsize{17.8} & \scriptsize{17.9} & \scriptsize{25.3} & \scriptsize{21.4} & \scriptsize{23.4} \\
\cline{2-15} 
 & \scriptsize{4} & -  & \scriptsize{2.31/74.23/5.59} & \scriptsize{2.37/79.18/4.42} &\scriptsize{21.7} & \scriptsize{15.9} & \scriptsize{15.9} & \scriptsize{12.8} & \scriptsize{14.7} & \scriptsize{13.6} & \scriptsize{13.2} & \scriptsize{16.4} & \scriptsize{13.9} & \scriptsize{15.2} \\
&  \scriptsize{4(a)}  & \scriptsize{$\mathcal{L}_{\text{ASR}}$}  & \scriptsize{2.53/79.68/6.39} & \scriptsize{2.58/83.47/5.51} & \scriptsize{17.0} & \scriptsize{15.5} & \scriptsize{13.2} & \scriptsize{11.8} & \scriptsize{11.7} & \scriptsize{11.0} & \scriptsize{11.4} & \scriptsize{13.3} & \scriptsize{12.4} & \scriptsize{12.9}\\
&  \scriptsize{4(b)} &  \scriptsize{$\mathcal{L}_{\text{ASR}}$ + $\mathcal{L}_{\text{SE}}$}& 
\scriptsize{2.38/76.36/5.77} & \scriptsize{2.42/80.81/4.60} & \scriptsize{18.5} & \scriptsize{15.8} & \scriptsize{14.4} & \scriptsize{12.4} & \scriptsize{12.6} & \scriptsize{12.1} & \scriptsize{12.2} & \scriptsize{14.4} & \scriptsize{13.1} & \scriptsize{13.8} \\
  \midrule[1.5pt]
\multirowcell{6}{\textbf{Sep. $\rightarrow$ Dervb.} \\ \scriptsize{(MVDR $\rightarrow$ DNN-WPE)}} & \scriptsize{5} & -  & \scriptsize{2.25/73.73/5.70} & \scriptsize{2.41/80.19/4.86} & \scriptsize{50.1} & \scriptsize{27.5} & \scriptsize{27.9} & \scriptsize{22.3} & \scriptsize{21.6} & \scriptsize{16.4} & \scriptsize{20.5} & \scriptsize{30.0} &\scriptsize{21.4} & \scriptsize{25.7} \\

& \scriptsize{5(a)} &  \scriptsize{$\mathcal{L}_{\text{ASR}}$} &
\scriptsize{2.46/78.62/6.46} & \scriptsize{2.58/82.96/5.80} & \scriptsize{39.9} & \scriptsize{27.4} & \scriptsize{20.4} & \scriptsize{18.9} & \scriptsize{16.2} & \scriptsize{17.7} & \scriptsize{15.8} & \scriptsize{23.1$^\dagger$} & \scriptsize{20.6} & \scriptsize{21.9$^\dagger$}\\
& \scriptsize{5(b)}  & \scriptsize{$\mathcal{L}_{\text{ASR}}$ + $\mathcal{L}_{\text{SE}}$} & 
\scriptsize{2.45/78.27/6.42} & \scriptsize{2.56/82.39/5.72} & \scriptsize{40.7} & \scriptsize{31.8} & \scriptsize{20.8} & \scriptsize{20.1} & \scriptsize{16.4} & \scriptsize{17.8} & \scriptsize{16.0} & \scriptsize{23.5$^\dagger$} & \scriptsize{22.1} & \scriptsize{22.8$^\dagger$}\\
\cline{2-15} 
 & \scriptsize{11} & -  & \scriptsize{2.36/76.90/5.88} & \scriptsize{2.46/81.71/5.04} & \scriptsize{21.1} & \scriptsize{15.7} & \scriptsize{15.2} & \scriptsize{12.5} & \scriptsize{14.1} & \scriptsize{12.1} & \scriptsize{13.5} &  \scriptsize{16.0} & \scriptsize{13.2} & \scriptsize{14.6}\\
 
&  \scriptsize{11(a)} &  \scriptsize{$\mathcal{L}_{\text{ASR}}$}  & \scriptsize{2.58/81.26/6.69} & \scriptsize{2.67/84.82/6.14} & \scriptsize{\textbf{16.7}} & \scriptsize{12.9} & \scriptsize{\textbf{12.8}} & \scriptsize{12.2} & \scriptsize{11.7} & \scriptsize{10.8} & \scriptsize{11.0} &  \scriptsize{\textbf{13.0}$^{\dagger}$} & \scriptsize{11.9$^{\dagger}$} & \scriptsize{12.5$^{\dagger}$}\\

& \scriptsize{11(b)} & \scriptsize{$\mathcal{L}_{\text{ASR}}$ + $\mathcal{L}_{\text{SE}}$} & \scriptsize{2.55/80.69/6.66} & \scriptsize{2.66/84.77/6.15} & \scriptsize{17.2} & \scriptsize{\textbf{12.0}} & \scriptsize{13.2} & \scriptsize{\textbf{11.6}} & \scriptsize{11.8} & \scriptsize{\textbf{10.3}} & \scriptsize{11.2} &  \scriptsize{13.3$^{\dagger \ast \ddagger}$} & \scriptsize{\textbf{11.2}$^{\dagger \ast \ddagger}$} & \scriptsize{\textbf{12.3}$^{\dagger \ast \ddagger}$}\\
\hline
\multirowcell{6}{\textbf{Sep. $\rightarrow$ Dervb.} \\ \scriptsize{(MVDR $\rightarrow$ SpecM)}} & \scriptsize{12} & -  & \scriptsize{2.35/76.60/7.37} & \scriptsize{2.48/80.75/6.62} & \scriptsize{52.7} & \scriptsize{31.2} & \scriptsize{30.7} & \scriptsize{25.3} & \scriptsize{23.6} & \scriptsize{19.6} & \scriptsize{22.5} & \scriptsize{32.4} & \scriptsize{24.6} & \scriptsize{28.5} \\

& \scriptsize{12(a)} &  \scriptsize{$\mathcal{L}_{\text{ASR}}$}  &\scriptsize{2.52/79.84/7.23} & \scriptsize{2.61/83.59/6.59} & \scriptsize{38.3} & \scriptsize{30.4} & \scriptsize{20.6} & \scriptsize{19.8} & \scriptsize{16.3} & \scriptsize{16.4} & \scriptsize{15.9} & \scriptsize{22.8$^{\dagger}$} & \scriptsize{21.2$^{\dagger}$} & \scriptsize{22.0$^{\dagger}$}\\
& \scriptsize{12(b)} & \scriptsize{$\mathcal{L}_{\text{ASR}}$ + $\mathcal{L}_{\text{SE}}$}& \scriptsize{2.49/79.16/6.61} & \scriptsize{2.56/82.99/5.91} & \scriptsize{39.4} & \scriptsize{30.3} & \scriptsize{20.9} & \scriptsize{19.4} & \scriptsize{16.2} & \scriptsize{17.1} & \scriptsize{16.2} & \scriptsize{23.2$^{\dagger}$} & \scriptsize{21.2$^{\dagger}$} & \scriptsize{22.2$^{\dagger}$}\\
\cline{2-15} 
 & \scriptsize{18} & -  & \scriptsize{2.51/81.03/\textbf{7.77}} & \scriptsize{2.58/82.99/\textbf{7.29}} & \scriptsize{22.0} & \scriptsize{17.2} & \scriptsize{16.8} & \scriptsize{13.4} & \scriptsize{14.5} & \scriptsize{13.1} & \scriptsize{14.4} & \scriptsize{16.9} & \scriptsize{14.2}& \scriptsize{15.6}\\

& \scriptsize{18(a)} &  \scriptsize{$\mathcal{L}_{\text{ASR}}$}  & \scriptsize{\textbf{2.60}/\textbf{81.64}/7.41} & \scriptsize{\textbf{2.68}/\textbf{85.22}/6.77} & \scriptsize{\textbf{16.7}} & \scriptsize{13.8} & \scriptsize{13.0} & \scriptsize{12.2} & \scriptsize{\textbf{11.6}} & \scriptsize{10.8} & \scriptsize{11.0} &  \scriptsize{13.1$^{\dagger}$} & \scriptsize{12.1$^{\dagger}$} & \scriptsize{12.6$^{\dagger}$}\\

&  \scriptsize{18(b)} & \scriptsize{$\mathcal{L}_{\text{ASR}}$ + $\mathcal{L}_{\text{SE}}$} & \scriptsize{2.55/80.65/6.70} & \scriptsize{2.60/84.43/6.05} & \scriptsize{\textbf{16.7}} & \scriptsize{13.4} & \scriptsize{13.0} & \scriptsize{12.3} & \scriptsize{11.9} & \scriptsize{10.6} & \scriptsize{\textbf{10.9}} &  \scriptsize{13.1$^{\dagger \ast \ddagger}$} & \scriptsize{11.9$^{\dagger \ast \ddagger}$} & \scriptsize{12.5$^{\dagger \ast \ddagger}$}\\
 \midrule[1.5pt]
\multirowcell{6}{ \textbf{Dervb. $\rightarrow$ Sep.} \\ \scriptsize{(DNN-WPE $\rightarrow$ MVDR)}} & \scriptsize{19} & -  & \scriptsize{2.46/79.75/6.44} & \scriptsize{2.67/84.68/6.32} & \scriptsize{47.2} & \scriptsize{25.4} & \scriptsize{24.6} & \scriptsize{15.6} & \scriptsize{19.2} & \scriptsize{13.2} & \scriptsize{19.2} & \scriptsize{27.5} & \scriptsize{17.1} & \scriptsize{22.6}\\

&  \scriptsize{19(a)} &  \scriptsize{$\mathcal{L}_{\text{ASR}}$}  &\scriptsize{2.61/81.91/6.86} & \scriptsize{2.70/85.21/6.28} & \scriptsize{37.8} & \scriptsize{22.2} & \scriptsize{18.8} & \scriptsize{17.2} & \scriptsize{14.9} & \scriptsize{13.1} & \scriptsize{15.0} & \scriptsize{21.6$^{\dagger}$} & \scriptsize{16.9} & \scriptsize{19.3$^{\dagger}$} \\

& \scriptsize{19(b)}  & \scriptsize{$\mathcal{L}_{\text{ASR}}$ + $\mathcal{L}_{\text{SE}}$} & \scriptsize{2.61/82.12/6.82} & \scriptsize{2.69/85.22/6.28} & \scriptsize{37.6} & \scriptsize{25.3} & \scriptsize{19.0} & \scriptsize{15.5} & \scriptsize{15.6} & \scriptsize{13.5} & \scriptsize{15.0} & \scriptsize{21.8$^{\dagger}$} & \scriptsize{17.2} & \scriptsize{19.5$^{\dagger}$}\\
\cline{2-15} 
 & \scriptsize{25} & -  & \scriptsize{2.56/82.61/6.61} & \scriptsize{2.72/85.85/6.49} & \scriptsize{20.8} & \scriptsize{15.0} & \scriptsize{15.1} & \scriptsize{10.9} & \scriptsize{12.4} & \scriptsize{10.7} & \scriptsize{12.3} & \scriptsize{15.1} & \scriptsize{11.8} & \scriptsize{13.5} \\
& \scriptsize{25(a)}  & \scriptsize{$\mathcal{L}_{\text{ASR}}$}  & \scriptsize{\textbf{2.71}/84.33/\textbf{7.04}} & \scriptsize{\textbf{2.75}/86.42/6.48} & \scriptsize{\textbf{16.0}} & \scriptsize{14.4} & \scriptsize{12.5} & \scriptsize{\textbf{10.6}} & \scriptsize{\textbf{10.7}} & \scriptsize{10.1} & \scriptsize{11.2} & \scriptsize{\textbf{12.6}$^{\dagger}$} & \scriptsize{11.4} & \scriptsize{12.0$^{\dagger}$} \\


& \scriptsize{25(b)}  &  \scriptsize{$\mathcal{L}_{\text{ASR}}$ + $\mathcal{L}_{\text{SE}}$} & 
\scriptsize{2.68/\textbf{84.75}/6.80} & \scriptsize{\textbf{2.75}/\textbf{86.82}/6.50} & \scriptsize{16.2} & \scriptsize{\textbf{13.3}} & \scriptsize{12.7} & \scriptsize{\textbf{10.6}} & \scriptsize{11.0} & \scriptsize{\textbf{9.9}} & \scriptsize{10.8} & \scriptsize{12.7$^{\dagger \ast \ddagger}$} & \scriptsize{\textbf{11.0}$^{\ast \ddagger}$} & \scriptsize{\textbf{11.9}$^{\dagger \ast \ddagger}$}\\

\hline
\multirowcell{6}{ \textbf{Dervb. $\rightarrow$ Sep.} \\ \scriptsize{(SpecM $\rightarrow$ MVDR)}} & \scriptsize{26} & -  & \scriptsize{2.24/73.72/6.54} & \scriptsize{2.51/80.67/6.32} & \scriptsize{57.0} & \scriptsize{30.4} & \scriptsize{32.8} & \scriptsize{20.4} & \scriptsize{25.6} & \scriptsize{14.9} & \scriptsize{24.2} & \scriptsize{34.9} & \scriptsize{20.8} & \scriptsize{27.9}\\

& \scriptsize{26(a)} &   \scriptsize{$\mathcal{L}_{\text{ASR}}$}  &\scriptsize{2.52/79.11/6.46} & \scriptsize{2.62/82.60/5.67} & \scriptsize{41.4} & \scriptsize{32.9} & \scriptsize{22.3} & \scriptsize{19.0} & \scriptsize{16.9} & \scriptsize{16.9} & \scriptsize{15.6} & \scriptsize{24.1$^{\dagger}$} & \scriptsize{21.7} & \scriptsize{22.9$^{\dagger}$}\\
& \scriptsize{26(b)}  & \scriptsize{$\mathcal{L}_{\text{ASR}}$ + $\mathcal{L}_{\text{SE}}$} & \scriptsize{2.53/79.57/6.50} & \scriptsize{2.65/83.12/5.91} & \scriptsize{42.5} & \scriptsize{29.8} & \scriptsize{21.7} & \scriptsize{19.9} & \scriptsize{16.7} & \scriptsize{16.5} & \scriptsize{16.1} & \scriptsize{24.3$^{\dagger}$} & \scriptsize{21.1} & \scriptsize{22.7$^{\dagger}$}\\
\cline{2-15} 
 & \scriptsize{32} & - & \scriptsize{2.36/77.10/6.76} & \scriptsize{2.57/82.20/\textbf{6.60}} & \scriptsize{24.9} & \scriptsize{15.5} & \scriptsize{16.9} & \scriptsize{13.2} & \scriptsize{15.3} & \scriptsize{11.2} & \scriptsize{14.8} &  \scriptsize{18.0} & \scriptsize{13.0} & \scriptsize{15.5}\\
 
& \scriptsize{32(a)} & \scriptsize{$\mathcal{L}_{\text{ASR}}$}  & \scriptsize{2.65/82.02/6.76} & \scriptsize{2.68/84.55/5.87} & \scriptsize{16.7} & \scriptsize{14.4} & \scriptsize{12.7} & \scriptsize{11.8} & \scriptsize{11.1} & \scriptsize{10.7} & \scriptsize{10.6} & \scriptsize{12.8$^{\dagger}$} & \scriptsize{12.1$^{\dagger}$} & \scriptsize{12.5$^{\dagger}$}\\

& \scriptsize{32(b)}  & \scriptsize{$\mathcal{L}_{\text{ASR}}$ + $\mathcal{L}_{\text{SE}}$} & \scriptsize{2.66/82.94/6.63} & \scriptsize{2.71/85.34/5.98} & \scriptsize{16.8} & \scriptsize{13.4} & \scriptsize{\textbf{12.4}} & \scriptsize{11.5} & \scriptsize{11.0} & \scriptsize{11.4} & \scriptsize{\textbf{10.3}} &  \scriptsize{\textbf{12.6}$^{\dagger \ast \ddagger}$} & \scriptsize{11.9$^{\dagger \ast \ddagger}$} & \scriptsize{12.3$^{\dagger \ast \ddagger}$}\\
 \midrule[1.5pt]
\multirowcell{6}{\textbf{Joint Sep. \& Dervb.} \\ \scriptsize{(WPD)}} & \scriptsize{33} & -  & \scriptsize{2.42/76.63/6.64} & \scriptsize{2.62/83.25/6.12} &  \scriptsize{55.0} & \scriptsize{28.8} & \scriptsize{30.1} & \scriptsize{17.4} & \scriptsize{22.6} & \scriptsize{15.0} & \scriptsize{21.8} & \scriptsize{32.4} & \scriptsize{19.4}& \scriptsize{25.9} \\

& \scriptsize{33(a)}  & \scriptsize{$\mathcal{L}_{\text{ASR}}$}& \scriptsize{2.52/78.55/6.97} & \scriptsize{2.63/82.88/6.18} &
\scriptsize{43.6} & \scriptsize{34.1} & \scriptsize{23.3} & \scriptsize{14.9} & \scriptsize{17.4} & \scriptsize{17.9} & \scriptsize{17.0} & \scriptsize{25.3$^{\dagger}$} & \scriptsize{20.8} & \scriptsize{23.1$^{\dagger}$}\\
& \scriptsize{33(b)} & \scriptsize{$\mathcal{L}_{\text{ASR}}$ + $\mathcal{L}_{\text{SE}}$} & \scriptsize{2.53/78.76/6.95} & \scriptsize{2.64/83.23/6.17} & \scriptsize{44.7} & \scriptsize{32.5} & \scriptsize{23.3} & \scriptsize{15.0} & \scriptsize{18.1} & \scriptsize{17.2} & \scriptsize{17.0} &  \scriptsize{25.7$^{\dagger}$} & \scriptsize{20.2} & \scriptsize{23.0$^{\dagger}$}\\

\cline{2-15} 
 & \scriptsize{36} & - & \scriptsize{2.53/79.64/6.80} & \scriptsize{2.67/84.45/6.29} &
 \scriptsize{24.6} & \scriptsize{15.8} & \scriptsize{16.3} & \scriptsize{12.1} & \scriptsize{13.7} & \scriptsize{11.3} & \scriptsize{13.6} & \scriptsize{17.0} & \scriptsize{12.7} & \scriptsize{14.9} \\
 
& \scriptsize{36(a)}  & \scriptsize{$\mathcal{L}_{\text{ASR}}$}  & \scriptsize{\textbf{2.61}/80.93/\textbf{7.16}} & \scriptsize{2.69/84.81/\textbf{6.39}}
&\scriptsize{\textbf{19.0}} & \scriptsize{\textbf{12.7}} & \scriptsize{\textbf{13.8}} & \scriptsize{11.0} & \scriptsize{\textbf{11.4}} & \scriptsize{\textbf{10.2}} & \scriptsize{\textbf{11.3}} &  \scriptsize{\textbf{13.9}$^{\dagger}$} & \scriptsize{\textbf{11.1}$^{\dagger}$} & \scriptsize{\textbf{12.5}$^{\dagger}$}\\

& \scriptsize{36(b)}  & \scriptsize{$\mathcal{L}_{\text{ASR}}$ + $\mathcal{L}_{\text{SE}}$} & \scriptsize{2.60/\textbf{81.27}/6.95} & \scriptsize{\textbf{2.70}/\textbf{85.15}/6.34} & \scriptsize{19.4} & \scriptsize{13.7} & \scriptsize{13.9} & \scriptsize{\textbf{10.5}} & \scriptsize{11.9} & \scriptsize{10.7} & \scriptsize{11.5} &  \scriptsize{14.2$^{\dagger \ast}$} & \scriptsize{11.4$^{\dagger \ast \ddagger}$} & \scriptsize{12.8$^{\dagger \ast \ddagger}$}\\
  \midrule[2pt]
\end{tabular}}
\vspace{-0.4cm}
\end{table*} 

\vspace{-0.4cm}
\subsection{Performance of Audio-visual Multi-channel Speech Enhancement and Recognition Systems} \label{subsection:pipelined_system}
In this part, we systematically investigate the performance improvements attributed to the visual modality in the proposed integrated speech enhancement architectures of Section \ref{section:AV_Sep&Dervb} on the LRS2 simulated multi-channel mixture dataset with four angle difference ranges [$0^{\circ}$, $15^{\circ}$), [$15^{\circ}$, $45^{\circ}$), [$45^{\circ}$, $90^{\circ}$) and [$90^{\circ}$, $180^{\circ}$).
The mask-based MVDR approach is used in the separation module, and the dereverberation module leverages either DNN-WPE or SpecM based dereverberation methods. The mask-based WPD is used for joint speech separation \& dereverberation. 
The multi-channel audio (including AF and IPD) features and visual modality features and their fusion mechanism presented in Sections \ref{subsection:audio_modality}, \ref{subsection:visual_modality}, \ref{subsection:modality_fusion} and \ref{subsection:av_dervb} for speech separation and dereverberation are used. 
The visual features are also incorporated into the Conformer speech recognition back-end, as described in Section \ref{section:comformer_asr}.
The speech recognition systems in Table \ref{tab:table4} are obtained by fine-tuning the baseline single-channel Conformer ASR (Table \ref{tab:table1}, sys. 1) or AVSR (Table \ref{tab:table1}, sys. 2) systems using the enhanced outputs of the corresponding speech enhancement front-ends.

From Table \ref{tab:table4}, several trends can be observed:

\textbf{1)} The proposed audio-visual multi-channel speech separation, dereverberation and recognition systems (sys. 11,18,25,32,36) consistently outperformed the corresponding audio-only baseline systems (sys. 5,12,19,26,33) on the LRS2 simulated test set. 
Consistent performance improvements in PESQ, STOI and SRMR scores were also obtained. 
For example, a statistically significant WER reduction of \textbf{12.4\% absolute} (\textbf{45.1\% relative}) was obtained by the full audio-visual system (sys. 25) over the corresponding audio-only baseline (sys. 19) using a pipelined front-end architecture whereby speech dereverberation was followed by separation.
A general trend can also be found that the performance gap between systems with full incorporation of video modality (sys. 11,18,25,32,36) and those using audio-only (sys. 5,12,19,26,33) was much larger when examining the performance on the more challenging subsets, e.g. when inter-speaker angle difference fell in the smallest range of [$0^{\circ}$, $15^{\circ}$).

\textbf{2)} When compared with audio-only dereverberation, incorporating visual information into the corresponding DNN-WPE (sys. 6,8,10,20,22,24 vs. sys. 5,7,9,19,21,23) or SpecM based dereverberation (sys. 13,15,17,27,29,31 vs. sys. 12,14,16,26,28,30) module produced consistent improvements in terms of PESQ, STOI and SRMR scores, irrespective of the underlying form of integration between speech separation and dereverberation components.  
A statistically significant WER reduction by up to \textbf{1.9\% absolute} (sys. 13 vs. sys. 12, \textbf{5.9\% relative}) was also obtained. 

\textbf{3)} Among the proposed architectures to integrate speech separation and dereverberation components within the speech enhancement front-end, a pipelined, full audio-visual configuration performing DNN-WPE based speech dereverberation followed by mask-based MVDR speech separation using visual input in both enhancement and recognition stages (sys. 25 vs. sys. 11,18,32,36) produced the lowest overall WERs. 

\textbf{4)} The integrated audio-visual speech separation, dereverberation and recognition systems (sys. 11,18,25,32,36) consistently outperformed the corresponding separation-only AVSR systems (sys. 4) in terms of PESQ, STOI and SRMR scores.
However, with regard to recognition performance, the SpecM based AVSR systems (sys. 18,32) and the mask-WPD based AVSR system (sys. 36) did not outperform the baseline system (sys. 4). 
The potential causes were: 
\textbf{a)} For systems using SpecM based dereverberation (sys. 18,32), although perceptually enhanced speech quality was obtained when compared to the corresponding baseline systems (sys. 4), the spectral artifacts caused by SpecM introduced a negative impact on downstream speech recognition performance; 
and \textbf{b)} For mask-based WPD systems, the number of filter taps and microphone channels together produced spatial-temporal PSD matrices in Eqns. (\ref{equation:wpd_speech_PSD})-(\ref{equation:wpd_time_varying_power_PSD}) larger than, for example, those in Eqns. (\ref{equation:mvdr_mask_PSD_1})-(\ref{equation:mvdr_mask_PSD_2}) for MVDR speech separation only, and thus increased difficulty in their inversion. 
This was further suggested by the larger variance flooring scaling $\varepsilon$=10$^{-4}$ in mask-based WPD than all the other systems shown in the ablation studies of Table \ref{tab:table3}. 
This issue can offset the benefit of joint speech separation \& dereverberation from WPD.

\textbf{5)} Finally, incorporating both the video modality and AF spatial features into the front-ends (e.g. sys. 3,10,17,24,31,35) consistently outperformed the comparable systems using either only AF features (sys. 1,5,12,19,26,33), or video features alone (sys. 2,8,15,22,29,34).

\subsection{Performance of End-to-end Joint Fine-tuning of Speech Enhancement Front-end and Recognition Back-end } \label{subsection:joint_system}
\begin{figure}[H]
    \centering
    \setlength{\abovecaptionskip}{0pt plus 1pt minus 3pt}
    \includegraphics[scale=0.35]{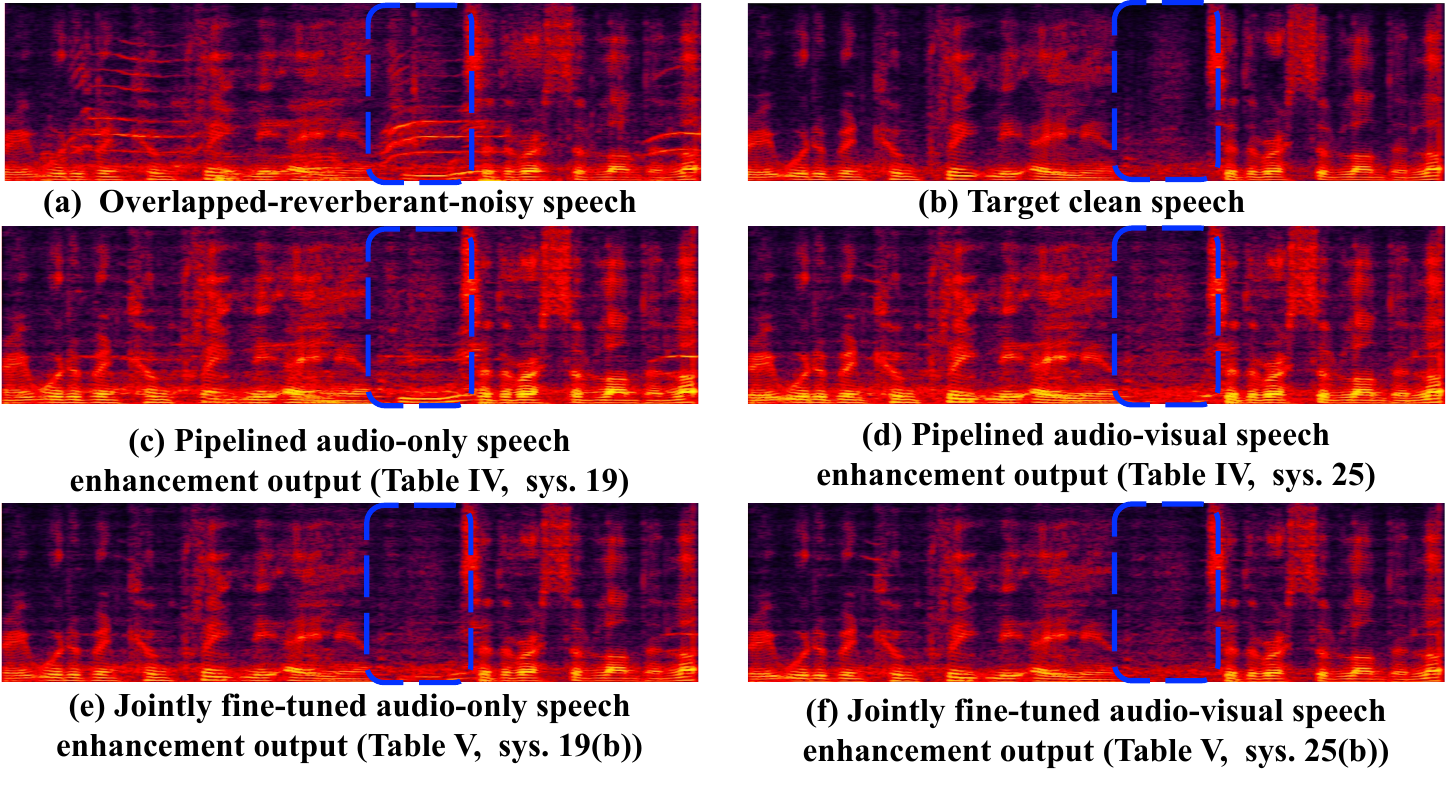}
    \caption{Example spectra of  \textbf{(a)} Overlapped-reverberant-noisy speech, \textbf{(b)} Target clean speech, \textbf{(c)} Pipelined audio-only speech enhancement output (Table~\ref{tab:table4}, sys. 19), \textbf{(d)} Pipelined audio-visual speech enhancement output (Table~\ref{tab:table4}, sys. 25), \textbf{(e)} Jointly fine-tuned audio-only speech enhancement output (Table~\ref{tab:table5}, sys. 19(b)), and \textbf{(f)} Jointly fine-tuned audio-visual speech enhancement output (Table~\ref{tab:table5}, sys. 25(b)). The spectrum portions circled using blue dotted lines in \textbf{(a)} represent the interfering speaker’s speech, background noise and reverberation, which have been largely removed in \textbf{(f)}.}
    \label{fig:spectra_module}
\vspace{-0.2cm}
\end{figure}

The most representative subset of audio-visual and audio-only multi-channel systems in Table \ref{tab:table4} are then end-to-end joint fine-tuning using either the ASR cost function alone, or a multi-task criterion interpolation between the speech enhancement and recognition cost as described in Section \ref{subsection:joint_fine-tuning}. 
Their performance in terms of WER and front-end metrics (PESQ, STOI and SRMR) are evaluated on both the LSR2 simulated (“Simu”) and replayed (“Replay”) test sets and shown in Table \ref{tab:table5} (original system numbering in Table \ref{tab:table4} carried over). Several main trends can be observed: 

\textbf{1)} After end-to-end joint fine-tuning, consistent performance improvements in WER were obtained over all systems without doing so (sys. marked with ``-" in Col. 3, Table \ref{tab:table5}), irrespective of the joint fine-tuning criterion based on ASR loss alone (sys. marked with ``(a)"), or its interpolation with enhancement loss (sys. marked with ``(b)").  In particular, statistically significant overall (``O.V.") WER reductions of \textbf{3.3\%} and \textbf{1.6\% absolute} (\textbf{14.6\%} and \textbf{11.9\% relative}) were obtained using the joint fine-tuned ASR (sys. 19(a) vs. sys. 19) and AVSR (sys. 25(b) vs. sys. 25) systems across both test sets. Consistent performance improvements in speech enhancement front-end metrics scores were also obtained. Fig.~\ref{fig:spectra_module} shows a set of example spectra of \textbf{(a)} Overlapped-reverberant-noisy speech, \textbf{(b)} Target clean speech, \textbf{(c)} Pipelined audio-only speech enhancement output (Table~\ref{tab:table4}, sys. 19), \textbf{(d)} Pipelined audio-visual speech enhancement output (Table~\ref{tab:table4}, sys. 25), \textbf{(e)} Jointly fine-tuned audio-only speech enhancement output (Table~\ref{tab:table5}, sys. 19(b)), and \textbf{(f)} Jointly fine-tuned audio-visual speech enhancement output (Table~\ref{tab:table5}, sys. 25(b)). The spectrum portions circled using blue dotted lines in \textbf{(a)} represent the interfering speaker’s speech, background noise and reverberation, which have been largely removed in \textbf{(f)}.


 \textbf{2)} The best overall performance was produced by the end-to-end joint fine-tuned audio-visual system with DNN-WPE based dereverberation followed by mask-based MVDR (sys.25(b)). 
Using this system statistically significant WER reductions of up to \textbf{9.1\%} and \textbf{6.2\% absolute} (\textbf{41.7\%} and \textbf{36.0\% relative}) were obtained on the LRS2 simulated and replayed test sets over the audio-only baseline (19(b)). 
 In addition, all the jointly fine-tuned audio-visual speech separation, dereverberation and recognition systems consistently outperformed the comparable baseline separation-only AVSR systems (e.g. sys. 11(b),18(b),25(b),32(b),36(b) vs. sys. 4(b)), with a statistically significant WER reduction up to \textbf{1.9\% absolute} (\textbf{13.8\% relative}) (sys. 25(b) vs. sys. 4(b)).

\textbf{3)} End-to-end joint fine-tuning of the speech enhancement front-end and recognition back-end is effective in mitigating the impact from spectral artifacts produced in SpecM based dereverberation \cite{kumar2022end} (e.g. sys. 12(b),18(b),26(b),32(b)). 
This leads to their smaller performance gap against systems using DNN-WPE dereverberation (sys. 5(b),11(b),19(b),25(b)), when compared the gap before joint fine-tuning. 

\textbf{4)} A further ablation study is conducted on the setting of the speech enhancement cost weight $\gamma$ in Eqn. (\ref{equation:interpolate_weight}) using three end-to-end joint fine-tuned multi-channel speech enhancement and recognition systems: sys. 1(b), 4(b) and 25(b) of Table \ref{tab:table5}. Their WER performance with respect to $\gamma$ on the LRS2 simulated (``Simu") and replayed (``Replay") test sets are shown in Table~\ref{tab:table6}. These results suggest that the performance of the audio-visual multi-channel speech separation, dereverberation and recognition system (sys. 25(b)) is largely insensitive to the setting of $\gamma \in [0, 0.75]$ during end-to-end joint fine-tuning using interpolated speech enhancement and ASR error costs. 
\begin{table}[H]
\vspace{-0.4cm}
\centering
\setlength{\abovecaptionskip}{0pt plus 1pt minus 4pt}
\caption{WER(\%) performance of end-to-end joint fine-tuned multi-channel speech enhancement and recognition systems 1({\rm b}), 4({\rm b}) and 25({\rm b}) of Table \ref{tab:table5} with respect to the speech enhancement cost weight $\gamma$ in Eqn. (\ref{equation:interpolate_weight}) on the LRS2 simulated (``Simu") and replayed (``Replay") test sets. 
}\label{tab:table6}
\resizebox{1.0\columnwidth}{!}{
\begin{tabular}{c|cc|cc|cc|cc|cc} 
 \toprule[1.5pt]
 \multirowcell{2}{\diagbox[width=4em]{Sys.}{$\gamma$}} & \multicolumn{2}{c|}{0} & \multicolumn{2}{c|}{0.25} & \multicolumn{2}{c|}{0.5} & \multicolumn{2}{c|}{0.75}  & \multicolumn{2}{c}{1}  \\
& \scriptsize{Simu} & \scriptsize{Replay} & \scriptsize{Simu} & \scriptsize{Replay} & \scriptsize{Simu} &  \scriptsize{Replay} & \scriptsize{Simu} & \scriptsize{Replay} &  \scriptsize{Simu} & \scriptsize{Replay}\\
\midrule[1pt]
\scriptsize{1(b)} & \scriptsize{24.0} & \scriptsize{22.8} & \scriptsize{24.7} & \scriptsize{22.4} & \scriptsize{25.3} & \scriptsize{21.4} & \scriptsize{27.2} & \scriptsize{22.7} & \scriptsize{31.3} & \scriptsize{23.4} \\
\hline
\scriptsize{4(b)} & \scriptsize{13.3} & \scriptsize{12.4} & \scriptsize{14.0} & \scriptsize{12.6} & \scriptsize{14.4} & \scriptsize{13.1} & \scriptsize{14.6} & \scriptsize{13.2} & \scriptsize{16.4} & \scriptsize{13.9} \\
\hline
\scriptsize{25(b)} & \scriptsize{\textbf{12.6}} & \scriptsize{\textbf{11.4}} & \scriptsize{\textbf{12.7}} & \scriptsize{\textbf{11.3}} & \scriptsize{\textbf{12.7}} & \scriptsize{\textbf{11.0}} & \scriptsize{\textbf{12.7}} & \scriptsize{\textbf{10.4}} & \scriptsize{15.1} & \scriptsize{11.8} \\
\bottomrule[1.5pt]
\end{tabular}}
\vspace{-0.2cm}
\end{table}

\textbf{5)} The performance of the most important systems shown in Table \ref{tab:table4} (sys. 1,4,5,11,12,18,19,25,26,32,33,36) and Table \ref{tab:table5} (sys. 1(b),4(b),5(b),11(b),12(b),18(b),19(b),25(b),26(b),32(b),
33(b),36(b)) are further evaluated on the LRS3 \cite{afouras2018lrs3} test set after applying the same multi-channel mixture speech simulation protocol of Algorithm \ref{alg:algorithm1}. These results are shown in Table VII. Similar trends of WER reductions and improvements on speech enhancement metric scores, as well as the same performance ranking among the corresponding systems previously shown in Table \ref{tab:table4} and Table \ref{tab:table5}, can also be found in Table \ref{tab:table7}.

\begin{table*}[!t]
\centering
\setlength{\abovecaptionskip}{0pt plus 1pt minus 4pt}
\caption{Performance of integrated architectures for audio-visual multi-channel speech separation (``Sep."), dereverberation (``Dervb.") and recognition (``RECG.") on the LRS3 test set simulated multi-channel mixture speech via the LRS2 data trained pipelined and jointly fine-tuned (using \textbf{$\mathcal{L}_{\text{ASR}}$ + $\mathcal{L}_{\text{SE}}$} cost function) systems in Table \ref{tab:table4} and Table \ref{tab:table5}, respectively.
``Arch.", ``AF", ``SpecM", ``Conf." and ``Avg."  denote the architecture, angle feature, spectral mapping, Conformer and average, respectively. ``$\dagger$", ``$\ast$" and ``${\ddag}$" denote a statistically significant WER difference over the ``Pipelined" systems, the corresponding audio-only baseline systems (sys. 5,12,19,26,33) in the ``Jointly fine-tuned" Column and separation-only AVSR baseline system (sys. 4) in the ``Jointly fine-tuned" Column, respectively.
}
\label{tab:table7}
\resizebox{1.6\columnwidth}{!}{
\begin{tabular}{c|c|c|c|cc|c|c|c} 
 \toprule[1.5pt]
 \multirow{3}{*}{\scriptsize Arch.} & \multirow{3}{*}{\scriptsize Sys.} & \multirow{3}{*}{\scriptsize +AF} &  \multicolumn{4}{c|}{\scriptsize +Visual Features} & \multicolumn{2}{c}{\scriptsize{PESQ$(\uparrow)$ / STOI$(\uparrow)$ / SRMR$(\uparrow)$ / WER$(\downarrow)$}}  \\
 \cline{4-9} 
 &  &  & \tiny Sep. &  \multicolumn{2}{c|}{\tiny Dervb.}  & \tiny Recg. & \multicolumn{2}{c}{\scriptsize{Avg.}}\\
& & & \tiny(MVDR) &\tiny (DNN-WPE) & \tiny (SpecM) & \tiny(Conf.) & \scriptsize Pipelined & \scriptsize Jointly fine-tuned \\
\midrule[1pt]
\multirowcell{2}{ \scriptsize{\textbf{Sep.}} \\ \tiny{(MVDR only)} } & 1 & \cmark & \xmark  & \multicolumn{2}{c|}{-} & \xmark  & \scriptsize{2.22/72.63/5.76/40.3} & \scriptsize{2.32/75.77/6.09/34.5}  \\
 
 & 4 & \cmark & \cmark  & \multicolumn{2}{c|}{-} & \cmark & \scriptsize{2.30/75.18/5.89/29.8} & \scriptsize{2.38/77.57/6.12/26.9}\\

 \midrule[1pt]
 \multirowcell{2}{\scriptsize{\textbf{Sep. $\rightarrow$ Dervb.}} \\ \tiny{(MVDR $\rightarrow$ DNN-WPE)} } & 5 & \cmark & \xmark  & \xmark & & \xmark 
 &\scriptsize{2.25/74.97/6.12/38.6} &\scriptsize{2.46/79.44/6.95/31.9$^{\dagger}$} \\

 & 11 & \cmark & \cmark  & \cmark & & \cmark & \scriptsize{2.34/77.71/6.28/29.5} & \scriptsize{2.55/81.64/7.24/25.1$^{\dagger\ast\ddag}$}  \\

 \cline{1-9}
 \multirowcell{2}{\scriptsize{\textbf{Sep. $\rightarrow$ Dervb.}} \\ \tiny{(MVDR $\rightarrow$ SpecM)} } & 12 & \cmark & \xmark & & \xmark & \xmark & \scriptsize{2.38/77.88/8.02/41.9} & \scriptsize{2.50/80.32/7.13/31.7$^{\dagger}$}\\
 & 18 & \cmark & \cmark & & \cmark & \cmark  & \scriptsize{2.51/81.14/8.46/31.1} & \scriptsize{2.56/81.81/7.17/25.3$^{\dagger\ast\ddag}$} \\
\midrule[1pt]
 \multirowcell{2}{\scriptsize{\textbf{Dervb. $\rightarrow$ Sep.}} \\ \tiny{(DNN-WPE $\rightarrow$ MVDR)} } & 19 & \cmark & \xmark  & \xmark & & \xmark  
 & \scriptsize{2.48/81.40/7.25/34.6} 
 & \scriptsize{2.66/83.88/\textbf{7.80}/28.9$^{\dagger}$} \\
 & 25 & \cmark & \cmark  & \cmark & &\cmark 
 & \scriptsize{2.55/83.22/7.32/27.2} 
 & \scriptsize{\textbf{2.69}/\textbf{85.73}/7.70/\textbf{23.9}$^{\dagger\ast\ddag}$} \\
 \cline{1-9}
 \multirowcell{2}{\scriptsize{\textbf{Dervb. $\rightarrow$ Sep.}} \\ \tiny{(SpecM $\rightarrow$ MVDR)} } & 26 & \cmark & \xmark  & &\xmark & \xmark 
 &\scriptsize{2.28/75.91/7.33/42.4}
 &\scriptsize{2.54/81.00/7.14/32.5$^{\dagger}$} \\

 & 32 & \cmark & \cmark  & &\cmark &\cmark &\scriptsize{2.35/77.73/7.37/32.2}
 &\scriptsize{2.61/83.16/7.20/25.6$^{\dagger\ast\ddag}$}  \\
\midrule[1pt]
 \multirowcell{2}{\scriptsize{\textbf{Joint Sep. \& Dervb.}} \\ \tiny{(WPD)}} & 33& \cmark & \multicolumn{3}{c|}{\xmark}  & \xmark & \scriptsize{2.45/78.84/7.27/39.4} & \scriptsize{2.54/80.46/7.46/34.1$^{\dagger}$}\\

 & 36 & \cmark& \multicolumn{3}{c|}{\cmark} & \cmark & \scriptsize{2.51/80.60/7.31/30.3} & \scriptsize{2.58/82.19/7.46/26.7$^{\dagger\ast}$}\\
\bottomrule[1.5pt]
\end{tabular}}
\vspace{-0.4cm}
\end{table*}

\section{Conclusion}
\vspace{-0.1cm}
In this paper, an audio-visual multi-channel speech separation, dereverberation and recognition approach featuring a full incorporation of visual information into all system components is proposed. The advantages of additional visual modality over using acoustic features only are demonstrated  consistently in mask-based MVDR speech separation, DNN-WPE or spectral mapping (SpecM) based speech dereverberation front-end and Conformer based ASR back-end. A set of audio-visual front-end architectures that integrates the speech separation and dereverberation modules in a pipelined or joint fashion are also derived. They are end-to-end jointly fine-tuned to minimize the error cost mismatch between the speech enhancement front-end and ASR back-end. 
Experiments were conducted on the mixture overlapped and reverberant speech data constructed using simulation or replay of the benchmark Oxford LRS2 dataset. 
The proposed audio-visual multi-channel speech separation, dereverberation and recognition systems consistently outperformed the comparable audio-only multi-channel baseline by 9.1\% and 6.2\% absolute (41.7\% and 36.0\% relative) in word error rate (WER) reductions, together with consistent improvements obtained on PESQ, STOI and SRMR based speech enhancement metrics. Future research will focus on improving system generalization to diverse microphone array geometrics and room acoustics.  

\section*{Acknowledgment}
This research is supported by Hong Kong RGC GRF grant No. 14200021, 14200218, 14200220, TRS T45-407/19N and Innovation \& Technology Fund grant No. ITS/218/21.
We would like to thank Wangyou Zhang for the insightful discussions in the preliminary experiments of WPD.

\bibliography{main.bib}
\bibliographystyle{IEEEtran.bst}

\end{document}